\documentclass[aps,prl,reprint,showpacs,longbibliography]{revtex4-1}
\pdfoutput=1
\usepackage{setspace}
\usepackage{amsfonts,amsmath,amssymb}
\usepackage{txfonts}
\usepackage{graphicx, color}
\usepackage[ugly]{units}

\begin{document}

\title{Fundamental Asymmetry in Quenches Between Integrable and Nonintegrable Systems}

\author{Marcos Rigol}
\affiliation{Department of Physics, The Pennsylvania State University, University Park, Pennsylvania 16802, USA}

\begin{abstract}
We study quantum quenches between integrable and nonintegrable hard-core boson models in the thermodynamic limit with numerical linked cluster expansions. We show that while quenches in which the initial state is a thermal equilibrium state of an integrable model and the final Hamiltonian is nonintegrable (quantum chaotic) lead to thermalization, the reverse is not true. While this might appear counterintuitive given the fact that the eigenstates of both Hamiltonians are related by a unitary transformation, we argue that it is generic. Hence, the lack of thermalization of integrable systems is robust against quenches starting from stationary states of nonintegrable ones. Nonintegrable systems thermalize independently of the nature of the initial Hamiltonian. 
\end{abstract}


\date{\today}

\maketitle

\paragraph{Introduction.}
The relaxation dynamics of observables in isolated many-body quantum systems far from equilibrium, as well as their description after relaxation, are topics that have generated much interest in recent years \cite{dalessio_kafri_15, polkovnikov_sengupta_review_11}. They are relevant to experiments with ultracold quantum gases \cite{greiner_mandel_02b, kinoshita_wenger_06, will_best_10, trotzky_chen_12, gring_kuhnert_12, will_iyer_15_97, langen_erne_15, schreiber_hodgman_15}. In the experiments, the (to-a-good-approximation) isolated gases are usually taken far from equilibrium by suddenly changing some parameter(s) (such as, e.g., suddenly changing the depth of an optical lattice \cite{greiner_mandel_02b, will_best_10, trotzky_chen_12, will_iyer_15_97}), a protocol known as a quench (or quantum quench, if the experimental system is in the quantum degeneracy regime). Computational studies of quenches have revealed that, so long as one is dealing with generic (chaotic) quantum many-body systems, observables relax to the predictions of traditional statistical mechanics (they thermalize) \cite{rigol_dunjko_08_34, rigol_09_39, rigol_09_43, eckstein_kollar_09, banuls_cirac_11, khatami_pupillo_13_84, sorg_vidmar_14}. This is a consequence of eigenstate thermalization \cite{rigol_dunjko_08_34, deutsch_91, srednicki_94, srednicki_99}. Evidence for thermalization in the thermodynamic limit has been obtained using numerical linked cluster expansions (NLCEs) \cite{rigol_14_90}, a technique also used here.

On the other hand, it has been shown that in quenches between integrable systems observables relax to stationary values that are not the ones predicted by traditional ensembles of statistical mechanics \cite{rigol_dunjko_07_27, cazalilla_06, rigol_muramatsu_06_26, kollar_eckstein_08, barthel_schollwock_08, iucci_cazalilla_09, cassidy_clark_11_55, calabrese_essler_11, cazalilla_iucci_12, caux_konik_12, gramsch_rigol_12_77, essler_evangelisti_12, collura_sotiriadis_13a, caux_essler_13, mussardo_13,  fagotti_essler_13a, wright_rigol_14_92, pozsgay_14a, pozsgay_14b,cardy_16}. Instead, generalized Gibbs ensembles need to be used to describe observables after relaxation \cite{rigol_dunjko_07_27}. A topic that has received less attention, and which is the focus of this work, is what happens if one quenches from a stationary state (e.g., a thermal state) of a nonintegrable system to an integrable one. These quenches are more challenging to study theoretically because the initial state cannot be obtained using the analytical or numerical techniques applicable at integrability. They are relevant to experiments as quenches to integrable systems are likely to start from stationary states of nonintegrable ones. If the initial states in those quenches were superpositions of (final) Hamiltonian eigenstates in which expectation values of observables are typical for the Haar measure \cite{tasaki_98, goldstein_lebowitz_06, popescu_06, eisert_friesdorf_review_15}, they would lead to thermalization. We implicitly show that this is not the case.

\paragraph{Numerical linked cluster expansions.}
To address what happens when one quenches from a nonintegrable system to an integrable one, and contrast it to the reverse quench, we use a recently introduced NLCE approach \cite{rigol_14_90}. Such an expansion, when converged, allows one to compute the infinite time average of an extensive observable $\hat{\mathcal{O}}$ per site after a quench in the thermodynamic limit \cite{wouters_denardis_14_93,rigol_14_95}: $\overline{\mathcal{O}(\tau)} = \overline{\text{Tr}[\hat{\rho}(\tau)\hat{\mathcal{O}}]} = \text{Tr}[\overline{\hat{\rho}(\tau)}\hat{\mathcal{O}}] \equiv \text{Tr}[\hat{\rho}^\text{DE}\hat{\mathcal{O}}]=\mathcal{O}^\text{DE}$, where $\overline{(\cdot)} = \text{lim}_{\tau'\rightarrow\infty} \nicefrac[]{1}{\tau'}\int_0^{\tau'} d\tau\,(\cdot)$ indicates the infinite time average, $\hat{\rho}(\tau)=\exp[-i\hat{H}\tau/\hbar]\,\hat{\rho}_\text{ini}\exp[i\hat{H}\tau/\hbar]$ is the density matrix of the time-evolving state ($\hat{\rho}_\text{ini}$ is the initial density matrix), and $\hat{\rho}^\text{DE} \equiv \overline{\hat{\rho}(\tau)}$ is the diagonal ensemble (DE) density matrix \cite{rigol_dunjko_08_34}. 

Within NLCEs, $O^\text{DE}\equiv \mathcal{O}^\text{DE}/L$ ($L$ is the number of lattice sites) is calculated in the thermodynamic limit as the sum over the contributions of all connected clusters $c$ on the lattice: $O^\text{DE}=\sum_{c}M(c)\times W_{O^\text{DE}}(c)$, where $M(c)$ is the number of ways per site in which cluster $c$ appears, and $W_{O^\text{DE}}(c)$ is its weight for $O^\text{DE}$. The latter is calculated using the inclusion-exclusion principle: $W_{O^\text{DE}}(c)=\mathcal{O}^\text{DE}(c)-\sum_{s \subset c} W_{O^\text{DE}}(s)$, where the sum runs over all connected subclusters of $c$, and $\mathcal{O}^\text{DE}(c)$ is the diagonal ensemble result for $\hat{\mathcal{O}}$ in cluster $c$. In practice, due to computational limitations, calculations are carried out in a finite number of clusters. Here, we use all maximally connected clusters with up to 18 sites \cite{rigol_14_90}. In what follows, whenever $O^\text{DE}$ is calculated from contributions of maximally connected clusters with up to $l$ sites, we report its value as $O^\text{DE}_l$.

One can also obtain the thermal equilibrium prediction after the quench, which would describe the system if it thermalizes, by carrying out the same calculation above using $\hat{\rho}^\text{GE} = \exp{[-(\hat{H}-\mu\hat{N})/T]}/Z$ instead of $\hat{\rho}^\text{DE}$, where $T$ is the temperature, $\mu$ is the chemical potential, $\hat{N}=\sum_i\hat{n}^{}_i$ is the total number of particle operator, and $Z=\text{Tr}\{\exp{[-(\hat{H}-\mu\hat{N})/T]}\}$ is the partition function (we set the Boltzmann constant to 1) \cite{rigol_bryant_06_25}. The temperature and the chemical potential are fixed so that the energy and number of particles per site are the same in the diagonal and grand canonical ensembles.

\paragraph{Model Hamiltonian and quenches.}
In this work, we study quenches in which the nonintegrable Hamiltonian $\hat{H}_\text{nonint}$ can be written as $\hat{H}_\text{nonint}(\Lambda)=\hat{H}_\text{int}+\Lambda\hat{W}$, where
{\setlength\arraycolsep{0.5pt}
\begin{eqnarray}
\hat{H}_\text{int}&=&\sum_i \left[-\hat{b}^\dagger_i \hat{b}^{}_{i+1} - 
\textrm{H.c.} + \left(\hat{n}^{}_i-\dfrac{1}{2}\right)\left( \hat{n}^{}_{i+1}-\dfrac{1}{2}\right)\right],\nonumber\\
\hat{W}&=& \sum_i\left[-\hat{b}^\dagger_i \hat{b}^{}_{i+2} - \textrm{H.c.}
+\left(\hat{n}^{}_i-\dfrac{1}{2}\right)\left( \hat{n}^{}_{i+2}-\dfrac{1}{2}\right)\right].
\label{eq:hamil}
\end{eqnarray}
$\hat{H}_\text{int}$ is an integrable Hamiltonian (the $XXZ$ model in the spin-1/2 language \cite{cazalilla_citro_review_11_63}), $\hat{W}$ is an integrability breaking operator, $\hat{b}^\dag_i\ (\hat{b}^{}_i)$ is the hard-core boson creation (annihilation) operator, $\hat{n}^{}_i=\hat{b}^\dag_i\hat{b}^{}_i$, and $\Lambda$ characterizes the departure from integrability.

The initial states considered are in thermal equilibrium, namely, their density matrix has the form $\hat{\rho}_\text{ini} = \exp{[-\hat{H}_\text{ini}/T_\text{ini}]}/Z_\text{ini}$, where $\hat{H}_\text{ini}$ can be $\hat{H}_\text{nonint}(\Lambda)$ or $\hat{H}_\text{int}$, and the final Hamiltonian is then $\hat{H}_\text{int}$ or $\hat{H}_\text{nonint}(\Lambda)$, respectively. We do calculations for different values of $T_\text{ini}$. By setting the initial chemical potential to zero we ensure that, since $\hat{H}_\text{int}$ and $\hat{H}_\text{nonint}(\Lambda)$ are particle-hole symmetric, the systems are always at half filling. Hence, to obtain the grand-canonical density matrix that would describe a system if it thermalizes after the quench, we only need to determine $T$. This is done by ensuring that the energy $E^\text{GE}_{18}=E^\text{DE}_{18}$, with a relative error smaller than $10^{-11}$. We note that quenches involving the $XXZ$ model have been extensively studied in recent years \cite{pozsgay_13, fagotti_collura_14, wouters_denardis_14_93, pozsgay_mestyan14, mierzejewski_prelovssek_14, rigol_14_95, goldstein_andrei_14, mierzejewski_prosen_15}. Findings in those studies motivated the discovery of new families of conserved quantities in the $XXZ$ model \cite{ilievski_medenjak_15, ilievski_denardis_15}.

\paragraph{Energies and temperatures after the quench.}
In order to gain an understanding of the effects of the quenches on the energy of the system, we compute the relative difference
\begin{equation}
 \Delta E_{18}(\Lambda)=\frac{\left|E^\text{DE}_{18}(\Lambda)-E^\text{GE}_{18}(\Lambda=0)\right|}{\left|E^\text{GE}_{18}(\Lambda=0)\right|}.
\end{equation}
For all results reported, $E^\text{DE}_{18}(\Lambda)$ and $E^\text{GE}_{18}(\Lambda)$ are converged to the thermodynamic limit result within machine precision \cite{rigol_14_90}. For quenches from the integrable to the nonintegrable Hamiltonian, $\Lambda$ characterizes the final Hamiltonian, while for quenches from the nonintegrable to the integrable Hamiltonian, $\Lambda$ characterizes the initial Hamiltonian. If $\Lambda$ is small, one expects that $E^\text{DE}(\Lambda)=E^\text{DE}(\Lambda=0)+(\partial E^\text{DE}/\partial\Lambda)|_{\Lambda=0}\,\Lambda+\text{O}(\Lambda^2)$, which means that $\Delta E(\Lambda)\propto\Lambda$ [$E^\text{DE}(\Lambda=0)=E^\text{GE}(\Lambda=0)$, no quench]. The results reported in Fig.~\ref{fig:energytemp} for quenches between integrable and nonintegrable Hamiltonians show that this is indeed the case. Remarkably, at the temperatures studied, the linear regime extends almost to $\Lambda=1$. We also calculated the temperature $T(\Lambda)$ in thermal equilibrium at the same energy per site as the diagonal ensemble (as described before). Results for $\Delta T(\Lambda)=|T(\Lambda)-T_\text{ini}|/T_\text{ini}$ vs $\Lambda$ are reported in the inset in Fig.~\ref{fig:energytemp}. We find that, in the quenches from the nonintegrable to the integrable Hamiltonian, $T<T_\text{ini}$ for the values of $T_\text{ini}$ considered, i.e., the quench ``cools'' the system (of course, this cannot remain true as $T_\text{ini}\rightarrow0$). For those quenches, one can see that $\Delta T\propto\Lambda$ almost all the way to $\Lambda=1$. On the other hand, in the quenches from the integrable to the nonintegrable Hamiltonian, we find that $T>T_\text{ini}$, and $\Delta T\propto\Lambda$ over a shorter range of values of $\lambda$.

\begin{figure}[!t]
 \includegraphics[width=0.77\linewidth]{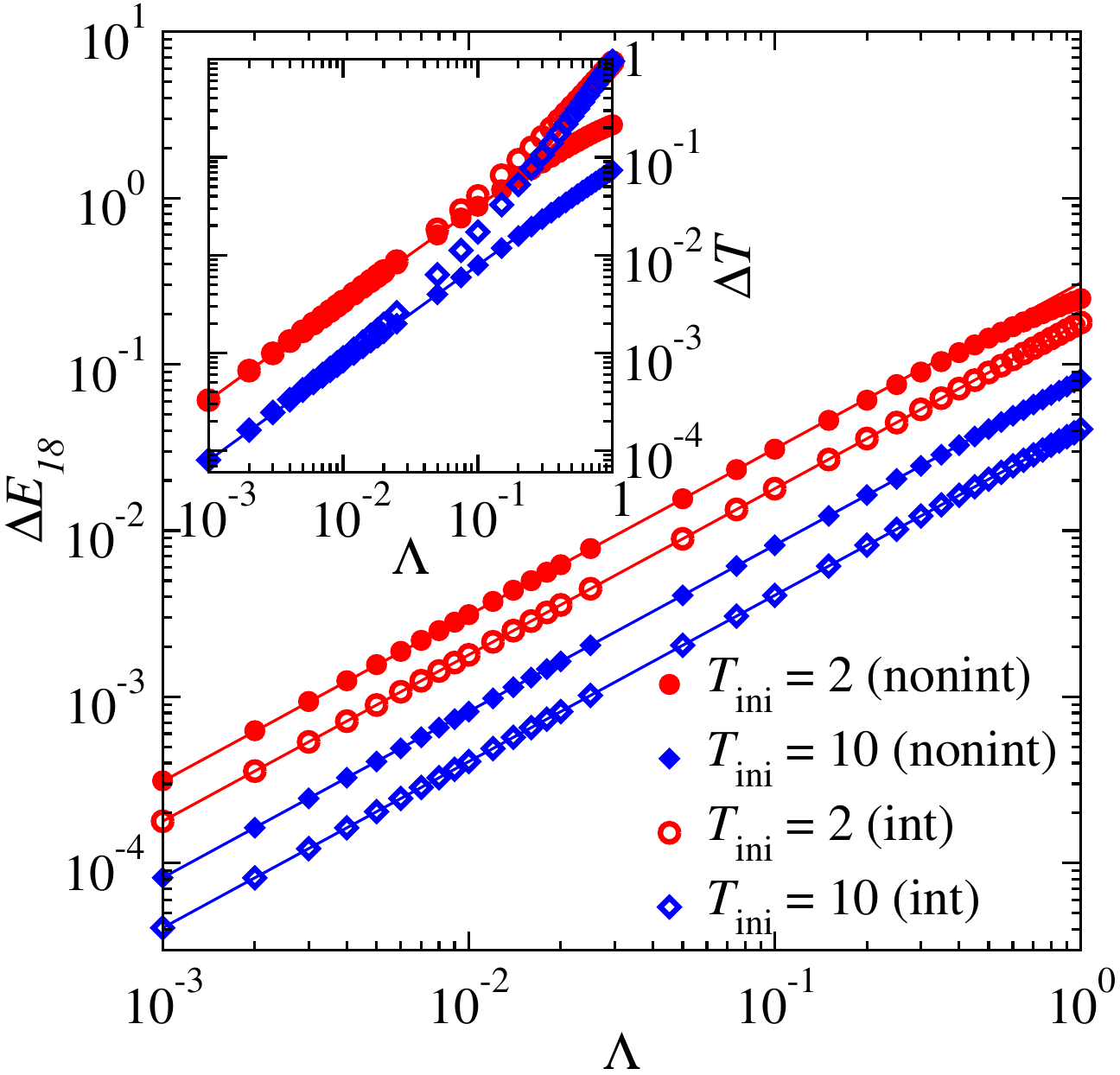} 
 \vspace{-0.25cm}
 \caption{The main panel shows the relative energy change $\Delta E_{18}$ from the equilibrium result for $\Lambda=0$ (see text) as a function of $\Lambda$, for two initial temperatures. Results are reported for quenches from initial thermal states of the integrable model to the nonintegrable one (open symbols) and from initial thermal states of the nonintegrable model to the integrable one (filled symbols). This convention will be used in the rest of the figures. The inset shows the relative change of the temperature (see text) for the same quenches as in the main panel. The lines (main panel and inset) report the results of a fit to $y=a\Lambda$, for $\Lambda=[10^{-3},10^{-1}]$.}
 \label{fig:energytemp}
\end{figure}

\paragraph{Diagonal vs thermal entropies.}
Having identified a weak quench regime in which energy and temperature changes in the {\it thermodynamic limit} are linear in $\Lambda$, we are ready to pose the first question we address in this work, namely, whether the energy distribution after a quench $W(E)=\sum_m |\rho^\text{DE}_{mm}|^2 \delta(E-E_m)$ ($\{E_m\}$ are the energy eigenvalues of the final Hamiltonian) from a nonintegrable to an integrable system is a smooth Gaussian function. This is expected to be the case if the eigenstates of the nonintegrable system, when decomposed in the eigenstates of the integrable one, lead to a (smooth) Gaussian energy distribution, as suggested by full exact diagonalization studies \cite{santos_borgonovi_12a, santos_borgonovi_12b, he_rigol_13_81}. NLCEs do not allow us to compute $W(E)$ directly. However, they do allow us to test whether it is a smooth function by computing the diagonal entropy $S^\text{DE}=-\text{Tr}[\hat{\rho}^\text{DE}\ln\hat{\rho}^\text{DE}]$, which is the appropriate entropy after a quench \cite{polkovnikov_11, santos_polkovnikov_11_57}. If $W(E)$ is a smooth Gaussian, then, in the thermodynamic limit, $S^\text{DE}=S^\text{GE}$ \cite{dalessio_kafri_15}. 

To see whether $S^\text{DE}=S^\text{GE}$ in the quenches studied here, we compute the following relative differences using NLCEs
\begin{equation}
 \delta S_l(\Lambda)=\frac{S^\text{GE}_{18}(\Lambda)-S^\text{DE}_l(\Lambda)}
 {S^\text{GE}_{18}(\Lambda)},
\end{equation}
where, again, $\Lambda$ characterizes the final Hamiltonian for quenches from integrable to nonintegrable systems, and the initial Hamiltonian for quenches from nonintegrable to integrable systems. For all results reported, $S^\text{GE}_{18}(\Lambda)$ is converged to the thermodynamic limit result within machine precision.

In Fig.~\ref{fig:entropy}(a), we plot $\delta S_l$ vs $l$ for six values of $\Lambda$ in the initial Hamiltonian (filled symbols). $\delta S_l(\Lambda=0)$ decreases exponentially to zero within machine precision [$S^\text{DE}_l(\Lambda=0)=S^\text{GE}_{l}(\Lambda=0)$, no quench]. However, as soon as $\Lambda\neq0$, one can see that $\delta S_l$ converges to a finite value as $l$ increases (the smaller the value of $\Lambda$, the smaller the quench, the smaller the value of $\delta S_l$ after convergence). This makes apparent that $S^\text{DE}$ after the quenches studied does not agree with $S^\text{GE}$. The former is always smaller. Hence, we conclude that $W(E)$ after quenches from nonintegrable to integrable systems is not a smooth Gaussian, and that the same must be true for the energy distribution of the eigenstates of the nonintegrable system when decomposed in the eigenstates of the integrable one. This is an instance, see Ref.~\cite{he_rigol_12_71} for another one, in which coarse graining  leads to the incorrect conclusion that $W(E)$ is smooth after a quench \cite{santos_borgonovi_12a, santos_borgonovi_12b, he_rigol_13_81}, and, hence, that $S^\text{DE}=S^\text{GE}$.

\begin{figure}[!t]
 \includegraphics[width=0.75\linewidth]{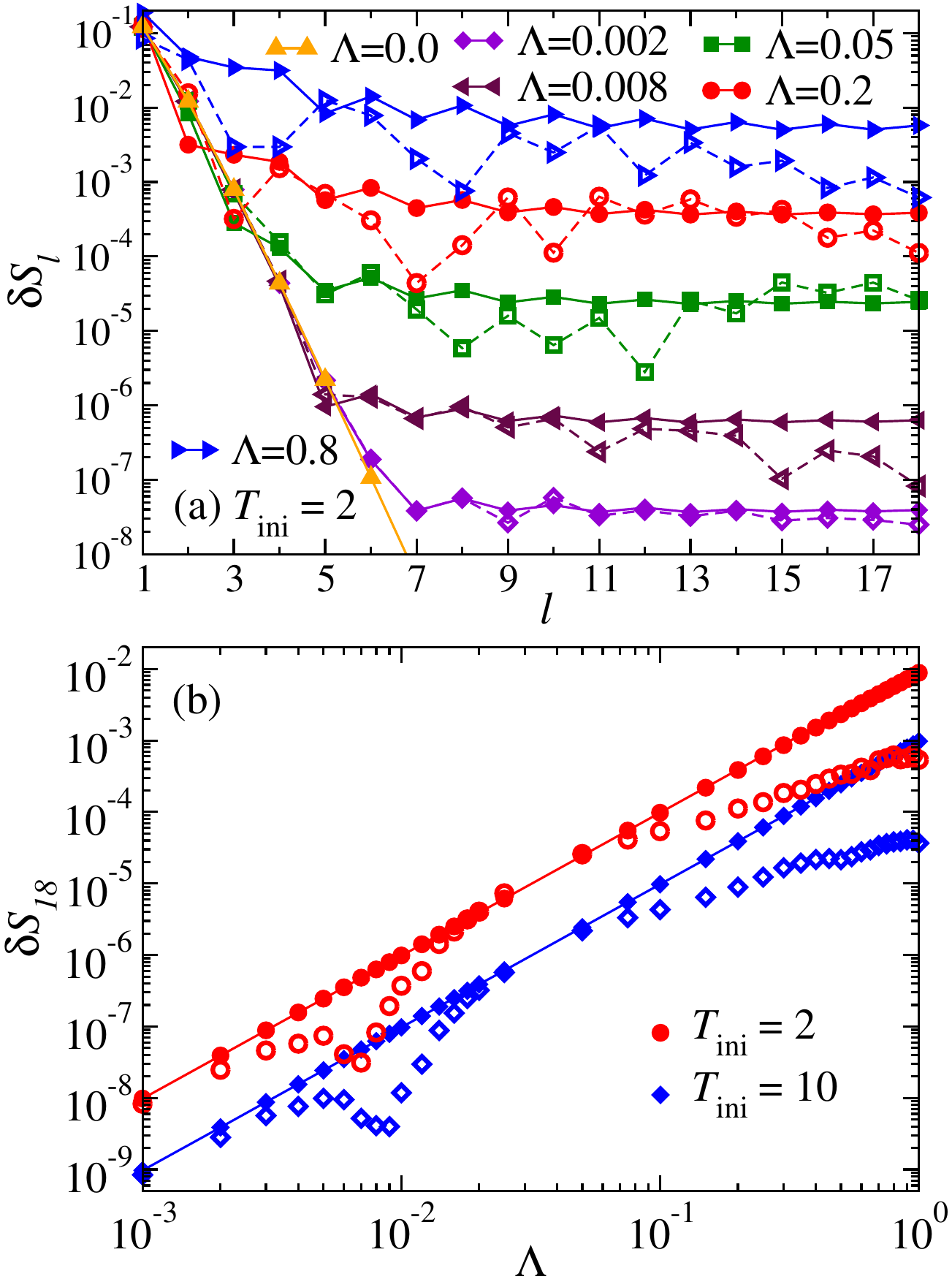} 
 \vspace{-0.25cm}
 \caption{(a) Relative entropy differences $\delta S_l$ (see text) vs $l$, for $T_\text{ini}=2$ and six values of $\Lambda$. $\delta S_l(\Lambda=0)$, i.e., no quench, decreases exponentially to zero within machine precision (it is $\sim$$10^{-12}$ for $l=10$). (b) $\delta S_{18}$ vs $\Lambda$ for two initial temperatures. The straight lines in (b) report the results of fits to $\delta S_{18}=a\Lambda^2$, for $\Lambda=[10^{-3},10^{-1}]$. As in Fig.~\ref{fig:energytemp}, results are reported for quenches from thermal states of the nonintegrable model to the integrable one (filled symbols) and vice versa (open symbols).}
 \label{fig:entropy}
\end{figure}

One might worry that the nonzero results in Fig.~\ref{fig:entropy}(a) are due to lack of NLCE convergence. However, in the weak quench regime, we can predict how $\delta S_l(\Lambda)$, if nonzero, should behave as a function of $\Lambda$ in the thermodynamic limit. For small values of $\Lambda$, we can expand $S^\text{DE}$ and $S^\text{GE}$ about $\Lambda=0$, i.e., $S^\text{DE/GE}(\Lambda)=S^\text{DE/GE}(\Lambda=0)+(\partial S^\text{DE/GE}/\partial\Lambda)|_{\Lambda=0}\,\Lambda+(\partial^2 S^\text{DE/GE}/\partial\Lambda^2)|_{\Lambda=0}\,\Lambda^2/2+\text{O}(\Lambda^3)$. As for the energy, $S^\text{DE}(\Lambda=0)=S^\text{GE}(\Lambda=0)$. Since $S^\text{GE}(\Lambda)\geq S^\text{DE}(\Lambda)$ no matter whether $\Lambda>0$ or $\Lambda<0$ [$S^\text{GE}(\Lambda)$ is the maximal entropy], we see that $(\partial S^\text{DE}/\partial\Lambda)|_{\Lambda=0}=(\partial S^\text{GE}/\partial\Lambda)|_{\Lambda=0}$ (there is no reason for them to vanish, we have checked that they do not) and that $\delta S_l(\Lambda)\propto \Lambda^2$. The fits to $\delta S_{18}(\Lambda)=a\Lambda^2$ in Fig.~\ref{fig:entropy}(b) show that this is the power law in our numerical data (independently of $T_\text{ini}$). These results strongly suggest $\delta S_{18}(\Lambda)$ is the thermodynamic limit result.

Next, we study the diagonal entropy in the reverse quenches, in which the initial state is a thermal equilibrium state of the integrable model and the Hamiltonian after the quench is nonintegrable. Figure~\ref{fig:entropy}(a) shows that in this case $\delta S_l(\Lambda)$ does not converge to a constant value. Instead, for the largest values of $l$ studied, $\delta S_l(\Lambda)$ can be seen to decrease with increasing $l$. $\delta S_{18}(\Lambda)$ vs $\Lambda$ is reported in Fig.~\ref{fig:entropy}(b) for two values of $T_\text{ini}$. One can see there that $\delta S_{18}(\Lambda)$ does not follow the expected $\delta S(\Lambda)=a\Lambda^2$ if nonvanishing. Actually, $\delta S_{18}(\Lambda)$ does not follow any power-law behavior at all, which suggests that it is the result of lack of convergence of the NLCE. We also note that, in quenches to nonintegrable systems, $\delta S_{18}(\Lambda)$ can decrease to values that are over an order of magnitude smaller than those in the quenches to integrable ones. Given these results, we expect $\delta S$ to vanish in the thermodynamic limit, no matter how small $\Lambda$ is after the quench. This means that in quenches from integrable to nonintegrable systems, $W(E)$ can be a smooth Gaussian.

\paragraph{Observables and thermalization.} 
If the answer to the first question posed previously would have been affirmative, it would have had an immediate consequence for experiments. It would have meant that observables thermalize after quenches from nonintegrable to integrable systems (the diagonal ensemble would predict the same expectation value of observables as thermal ensembles). Having concluded that the answer is negative, there is still the possibility that observables thermalize after the aforementioned quenches. In Ref. \cite{rigol_srednicki_12_69} it was argued that if, as we do here, one starts the quench from a stationary state of a nonintegrable system whose Hamiltonian $\hat{H}_\text{nonint}(\Lambda)=\hat{H}_\text{int}+\Lambda\hat{W}$, and quenches to the integrable limit, thermalization would occur. The idea is that the initial states constructed this way would provide an unbiased sampling of the eigenstates of the integrable model (within the microcanonical window) and, hence, would lead to thermalization after the system dephased. Numerical evidence that this is the case was obtained using full exact diagonalization in Refs.~\cite{rigol_srednicki_12_69,he_rigol_13_81}, but only for small quenches, namely, when the nonintegrable system was close to the integrable one. 

\begin{figure*}[!t]
 \includegraphics[width=0.8\linewidth]{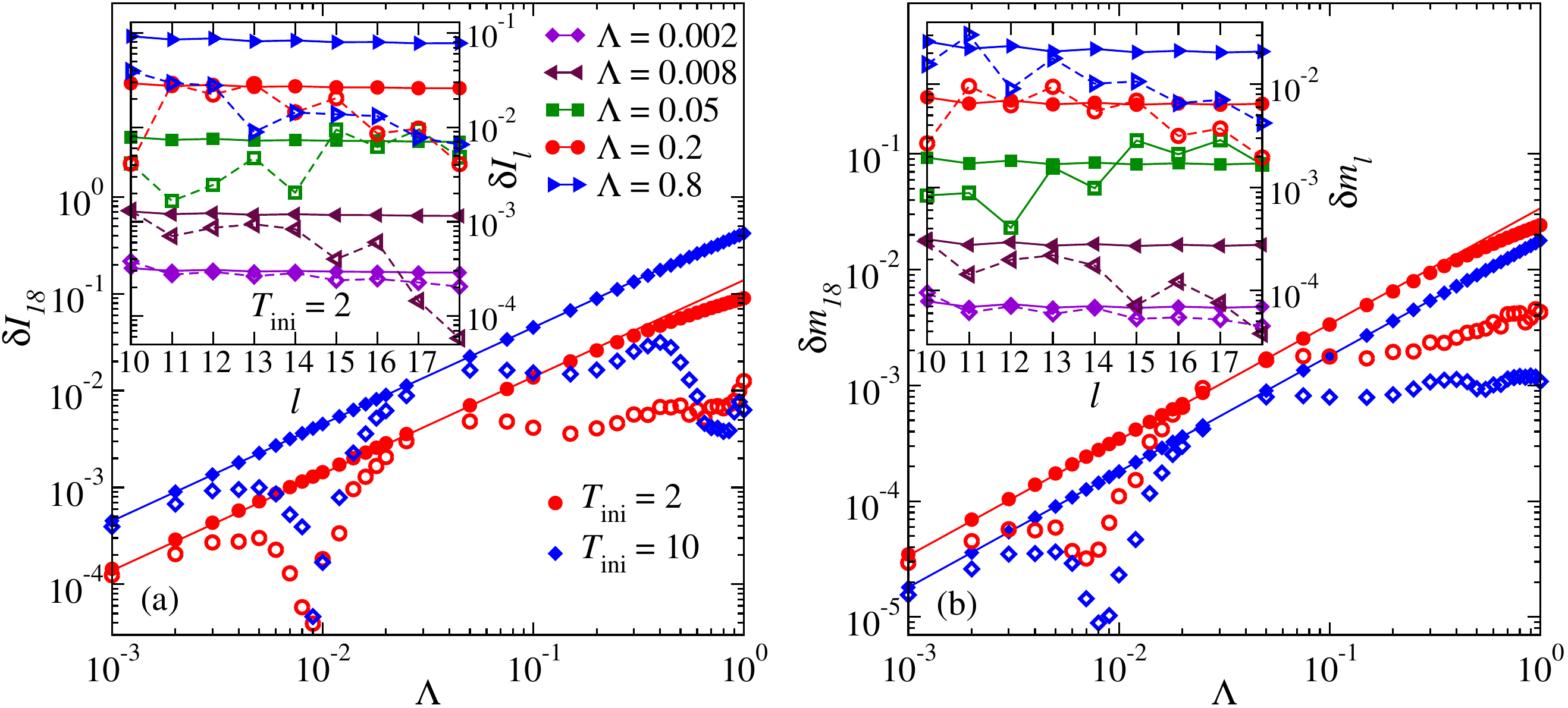} 
 \vspace{-0.2cm}
 \caption{The main panels show $\delta I_{18}$ (see text) vs $\Lambda$ (a), and $\delta m_{18}$ (see text) vs $\Lambda$ (b), for two initial temperatures. The straight lines in both panels report the results of a fit to $y=a\Lambda$, for $\Lambda=[10^{-3},10^{-1}]$. The insets show $\delta I_{18}$ vs $l$ (a), and $\delta m_{18}$ vs $l$ (b), for $T_\text{ini}=2$ and five values of $\Lambda$ [reported to the right of the inset in (a)]. As in Figs.~\ref{fig:energytemp} and \ref{fig:entropy}, results are reported for quenches from thermal states of the nonintegrable model to the integrable one (filled symbols) and vice versa (open symbols).}
 \label{fig:observables}
\end{figure*}

To check whether in the thermodynamic limit thermalization occurs in quenches from nonintegrable to integrable systems, we focus on two observables. The first one is local, the interaction energy per site $\hat{I}$, and the second one is nonlocal, the momentum distribution function $\hat{m}_k$:
\begin{eqnarray}
 \hat{I}&=&\frac1L\sum_i\left[ \left(\hat{n}^{}_i-\dfrac{1}{2}\right)\left(\hat{n}^{}_{i+1}-\dfrac{1}{2}\right) + \Lambda \left(\hat{n}^{}_i-\dfrac{1}{2}\right)\left(\hat{n}^{}_{i+2}-\dfrac{1}{2}\right) \right], \nonumber\\
 \hat{m}_k &=& \frac1L\sum_{jl}e^{ik(j-l)}\hat{b}_j^\dag \hat{b}^{}_{l}.
\end{eqnarray}
Those two observables can be measured in experiments with ultracold atoms \cite{cazalilla_citro_review_11_63, bloch_dalibard_review_08}. As for the entropy, we compute the following relative differences
\begin{eqnarray}
 \delta I_l(\Lambda)&=&\frac{\left|I^\text{DE}_{l}(\Lambda)-I^\text{GE}_{18}(\Lambda)\right|}
 {|I^\text{GE}_{18}(\Lambda)|},\nonumber\\
 \delta m_l(\Lambda)&=&\frac{\sum_k\left|{m_k}^\text{DE}_{l}(\Lambda)-{m_k}^\text{GE}_{18}(\Lambda)\right|}
 {\sum_k {m_k}^\text{GE}_{18}(\Lambda)},
\end{eqnarray}
and we have ensured that $I^\text{GE}_{18}(\Lambda)$ and ${m_k}^\text{GE}_{18}(\Lambda)$ are converged within machine precision. This constraint on ${m_k}^\text{GE}_{18}(\Lambda)$ is what restricts the initial temperatures to values $T_\text{ini}>1$. For other observables, the grand canonical ensemble results converge within machine precision for smaller values of $T_\text{ini}$. However, as made apparent in the figures, whenever converged our results are qualitatively similar independently of $T_\text{ini}$.

In the insets in Fig.~\ref{fig:observables}, we show $\delta I_l(\Lambda)$ (a), and $\delta m_l(\Lambda)$ (b), vs $l$ for $T_\text{ini}=2$ and five values of $\Lambda$ for quenches from nonintegrable to integrable systems (filled symbols). As for $\delta S_l(\Lambda)$ vs $l$ in Fig.~\ref{fig:entropy}, we find that $\delta I_l(\Lambda)$ and $\delta m_l(\Lambda)$ converge to nonzero results with increasing $l$. Hence, those observables are not described by thermal ensembles in the thermodynamic limit, i.e., they do not thermalize. Results for $\delta I_{18}(\Lambda)$ and $\delta m_{18}(\Lambda)$ vs $\Lambda$, for two initial temperatures, are reported in the main panels in Fig.~\ref{fig:observables}. A fit to the data in the interval $\Lambda=[10^{-3},10^{-1}]$ reveals that $\delta I_{18}(\Lambda)$ and $\delta m_{18}(\Lambda)$ are linear in $\Lambda$, as $\Delta E_{18}(\Lambda)$ was shown to be in Fig.~\ref{fig:energytemp}. This can be understood using the same analysis as for $\delta S_l$, but noting that the thermal predictions for $\hat{I}$ and $\hat{m}_k$ are not maximal as for the entropy. These results strongly suggest that, as for the entropy, the nonzero values of $\delta I_{18}(\Lambda)$ and $\delta m_{18}(\Lambda)$ have converged to the thermodynamic limit results.

On the other hand, for quenches from the integrable to the nonintegrable systems, the results in the insets (open symbols in Fig.~\ref{fig:observables}) decrease with increasing $l$, which suggest that $\delta I_l(\Lambda)$ and $\delta m_l(\Lambda)$ vanish in the thermodynamic limit. Like in the quenches studied in Ref.~\cite{rigol_14_90}, in the insets in Fig.~\ref{fig:observables} one can see a nearly linear decrease of $\delta I_l(\Lambda)$ and $\delta m_l(\Lambda)$ vs $l$ when $\Lambda>0.05$. This suggests that the approach to zero is exponential. In Ref.~\cite{iyer_srednicki_15_103}, it was argued that such an exponential convergence with $l$ is generic in numerical linked cluster expansions in unordered phases. The erratic behavior of $\delta I_{18}(\Lambda)$ and $\delta m_{18}(\Lambda)$ vs $\Lambda$ (open symbols in the main panels in Fig.~\ref{fig:observables}), and the fact that $\delta I_{18}(\Lambda)$ and $\delta m_{18}(\Lambda)$ are sometimes orders of magnitude smaller than the results for the reverse quench, suggest that the former are nonzero only because the NLCE results have not converged to the thermodynamic limit results [$\delta I(\Lambda)=0$ and $\delta m(\Lambda)=0$] within machine precision.

\paragraph{Summary and discussion.} 
In summary, we have shown that there is a fundamental asymmetry between quenches from nonintegrable to integrable systems and quenches from integrable to nonintegrable systems. The former do not lead to smooth Gaussian energy densities (as revealed by the diagonal entropy, which is different from the thermal one) and do not lead to thermalization (a consequence of the exponential sparseness of the energy densities). On the other hand, quenches to nonintegrable systems produce diagonal entropies that are thermal and lead to thermalization \cite{santos_polkovnikov_11_57,rigol_14_90}. This asymmetry might appear counterintuitive considering that  the bases of eigenstates of integrable and nonintegrable Hamiltonians are related by unitary transformations. If an eigenstate of the integrable Hamiltonian exhibits a smooth Gaussian energy density when written in terms of the eigenstates of the nonintegrable one, why would the reverse not be true (our conclusion here)? A way to justify this asymmetry is provided by the fact that integrable systems have an infinite number of local and extensive conserved quantities. For an eigenstate of a nonintegrable Hamiltonian (or, for that matter, any arbitrary initial state) to lead to a Gaussian energy density when written in terms of the eigenstates of the integrable one, it would need to have the same expectation value of all conserved quantities as the integrable system in thermal equilibrium (the energy of the latter is determined by the expectation value of the integrable Hamiltonian in that state). More precisely, in the Bethe-ansatz language in the thermodynamic limit, one would need the Bethe root densities to be those in thermal equilibrium \cite{caux_essler_13}. There is no reason why this should happen in general. It can happen if the integrability breaking perturbation is infinitesimally small so that the eigenstate of the nonintegrable Hamiltonian is a superposition of eigenstates of the integrable one within the microcanonical shell with the same energy. This explains why evidence for thermalization was seen in Refs. \cite{rigol_srednicki_12_69, he_rigol_13_81} {\it only} for small integrability breaking perturbations. Our results suggest that lack of thermalization will be robust in experimental studies of quenches to integrability.

\begin{acknowledgments}
This work was supported by the U.S. Office of Naval Research. The computations were performed in the Institute for CyberScience at Penn State. The visit of the author to KITP, where part of the letter was written, was supported by the National Science Foundation under Grant No.~NSF PHY11-25915. The author thanks M. Fagotti, S. Goldstein, T. Hara, D. Huse, P. Reimann, and H. Tasaki for motivating discussions, and R. Mondaini, A. Polkovnikov, L. F. Santos, and L. Vidmar for comments on the Letter. 
\end{acknowledgments}

\bibliography{paper}

\begin{thebibliography}{66}%
\makeatletter
\providecommand \@ifxundefined [1]{%
 \@ifx{#1\undefined}
}%
\providecommand \@ifnum [1]{%
 \ifnum #1\expandafter \@firstoftwo
 \else \expandafter \@secondoftwo
 \fi
}%
\providecommand \@ifx [1]{%
 \ifx #1\expandafter \@firstoftwo
 \else \expandafter \@secondoftwo
 \fi
}%
\providecommand \natexlab [1]{#1}%
\providecommand \enquote  [1]{``#1''}%
\providecommand \bibnamefont  [1]{#1}%
\providecommand \bibfnamefont [1]{#1}%
\providecommand \citenamefont [1]{#1}%
\providecommand \href@noop [0]{\@secondoftwo}%
\providecommand \href [0]{\begingroup \@sanitize@url \@href}%
\providecommand \@href[1]{\@@startlink{#1}\@@href}%
\providecommand \@@href[1]{\endgroup#1\@@endlink}%
\providecommand \@sanitize@url [0]{\catcode `\\12\catcode `\$12\catcode
  `\&12\catcode `\#12\catcode `\^12\catcode `\_12\catcode `\%12\relax}%
\providecommand \@@startlink[1]{}%
\providecommand \@@endlink[0]{}%
\providecommand \url  [0]{\begingroup\@sanitize@url \@url }%
\providecommand \@url [1]{\endgroup\@href {#1}{\urlprefix }}%
\providecommand \urlprefix  [0]{URL }%
\providecommand \Eprint [0]{\href }%
\providecommand \doibase [0]{http://dx.doi.org/}%
\providecommand \selectlanguage [0]{\@gobble}%
\providecommand \bibinfo  [0]{\@secondoftwo}%
\providecommand \bibfield  [0]{\@secondoftwo}%
\providecommand \translation [1]{[#1]}%
\providecommand \BibitemOpen [0]{}%
\providecommand \bibitemStop [0]{}%
\providecommand \bibitemNoStop [0]{.\EOS\space}%
\providecommand \EOS [0]{\spacefactor3000\relax}%
\providecommand \BibitemShut  [1]{\csname bibitem#1\endcsname}%
\let\auto@bib@innerbib\@empty
\bibitem [{\citenamefont {D'Alessio}\ \emph {et~al.}()\citenamefont
  {D'Alessio}, \citenamefont {Kafri}, \citenamefont {Polkovnikov},\ and\
  \citenamefont {Rigol}}]{dalessio_kafri_15}%
  \BibitemOpen
  \bibfield  {author} {\bibinfo {author} {\bibfnamefont {L.}~\bibnamefont
  {D'Alessio}}, \bibinfo {author} {\bibfnamefont {Y.}~\bibnamefont {Kafri}},
  \bibinfo {author} {\bibfnamefont {A.}~\bibnamefont {Polkovnikov}}, \ and\
  \bibinfo {author} {\bibfnamefont {M.}~\bibnamefont {Rigol}},\ }\bibfield
  {title} {\enquote {\bibinfo {title} {From quantum chaos and eigenstate
  thermalization to statistical mechanics and thermodynamics},}\ }\href@noop {}
  {\bibinfo  {journal} {arXiv:1509.06411}\ }\BibitemShut {NoStop}%
\bibitem [{\citenamefont {Polkovnikov}\ \emph {et~al.}(2011)\citenamefont
  {Polkovnikov}, \citenamefont {Sengupta}, \citenamefont {Silva},\ and\
  \citenamefont {Vengalattore}}]{polkovnikov_sengupta_review_11}%
  \BibitemOpen
\bibfield  {journal} {  }\bibfield  {author} {\bibinfo {author} {\bibfnamefont
  {A.}~\bibnamefont {Polkovnikov}}, \bibinfo {author} {\bibfnamefont
  {K.}~\bibnamefont {Sengupta}}, \bibinfo {author} {\bibfnamefont
  {A.}~\bibnamefont {Silva}}, \ and\ \bibinfo {author} {\bibfnamefont
  {M.}~\bibnamefont {Vengalattore}},\ }\bibfield  {title} {\enquote {\bibinfo
  {title} {\textit{Colloquium}: Nonequilibrium dynamics of closed interacting
  quantum systems},}\ }\href {\doibase 10.1103/RevModPhys.83.863} {\bibfield
  {journal} {\bibinfo  {journal} {Rev. Mod. Phys.}\ }\textbf {\bibinfo {volume}
  {83}},\ \bibinfo {pages} {863--883} (\bibinfo {year} {2011})}\BibitemShut
  {NoStop}%
\bibitem [{\citenamefont {Greiner}\ \emph {et~al.}(2002)\citenamefont
  {Greiner}, \citenamefont {Mandel}, \citenamefont {H\"ansch},\ and\
  \citenamefont {Bloch}}]{greiner_mandel_02b}%
  \BibitemOpen
  \bibfield  {author} {\bibinfo {author} {\bibfnamefont {M.}~\bibnamefont
  {Greiner}}, \bibinfo {author} {\bibfnamefont {O.}~\bibnamefont {Mandel}},
  \bibinfo {author} {\bibfnamefont {T.~W.}\ \bibnamefont {H\"ansch}}, \ and\
  \bibinfo {author} {\bibfnamefont {I.}~\bibnamefont {Bloch}},\ }\bibfield
  {title} {\enquote {\bibinfo {title} {Collapse and revival of the matter wave
  field of a {Bose-Einstein} condensate},}\ }\href@noop {} {\bibfield
  {journal} {\bibinfo  {journal} {Nature}\ }\textbf {\bibinfo {volume} {419}},\
  \bibinfo {pages} {51--54} (\bibinfo {year} {2002})}\BibitemShut {NoStop}%
\bibitem [{\citenamefont {Kinoshita}\ \emph {et~al.}(2006)\citenamefont
  {Kinoshita}, \citenamefont {Wenger},\ and\ \citenamefont
  {Weiss}}]{kinoshita_wenger_06}%
  \BibitemOpen
  \bibfield  {author} {\bibinfo {author} {\bibfnamefont {T.}~\bibnamefont
  {Kinoshita}}, \bibinfo {author} {\bibfnamefont {T.}~\bibnamefont {Wenger}}, \
  and\ \bibinfo {author} {\bibfnamefont {D.~S.}\ \bibnamefont {Weiss}},\
  }\bibfield  {title} {\enquote {\bibinfo {title} {A quantum {Newton's}
  cradle},}\ }\href@noop {} {\bibfield  {journal} {\bibinfo  {journal}
  {Nature}\ }\textbf {\bibinfo {volume} {440}},\ \bibinfo {pages} {900}
  (\bibinfo {year} {2006})}\BibitemShut {NoStop}%
\bibitem [{\citenamefont {Will}\ \emph {et~al.}(2010)\citenamefont {Will},
  \citenamefont {Best}, \citenamefont {Schneider}, \citenamefont
  {Hackerm\"uller}, \citenamefont {L\"uhmann},\ and\ \citenamefont
  {Bloch}}]{will_best_10}%
  \BibitemOpen
  \bibfield  {author} {\bibinfo {author} {\bibfnamefont {S.}~\bibnamefont
  {Will}}, \bibinfo {author} {\bibfnamefont {T.}~\bibnamefont {Best}}, \bibinfo
  {author} {\bibfnamefont {U.}~\bibnamefont {Schneider}}, \bibinfo {author}
  {\bibfnamefont {L.}~\bibnamefont {Hackerm\"uller}}, \bibinfo {author}
  {\bibfnamefont {D.-S.}\ \bibnamefont {L\"uhmann}}, \ and\ \bibinfo {author}
  {\bibfnamefont {I.}~\bibnamefont {Bloch}},\ }\bibfield  {title} {\enquote
  {\bibinfo {title} {Time-resolved observation of coherent multi-body
  interactions in quantum phase revivals},}\ }\href@noop {} {\bibfield
  {journal} {\bibinfo  {journal} {Nature}\ }\textbf {\bibinfo {volume} {465}},\
  \bibinfo {pages} {197--201} (\bibinfo {year} {2010})}\BibitemShut {NoStop}%
\bibitem [{\citenamefont {Trotzky}\ \emph {et~al.}(2012)\citenamefont
  {Trotzky}, \citenamefont {Chen}, \citenamefont {Flesch}, \citenamefont
  {McCulloch}, \citenamefont {Schollw\"ock}, \citenamefont {Eisert},\ and\
  \citenamefont {Bloch}}]{trotzky_chen_12}%
  \BibitemOpen
  \bibfield  {author} {\bibinfo {author} {\bibfnamefont {S.}~\bibnamefont
  {Trotzky}}, \bibinfo {author} {\bibfnamefont {Y.-A.}\ \bibnamefont {Chen}},
  \bibinfo {author} {\bibfnamefont {A.}~\bibnamefont {Flesch}}, \bibinfo
  {author} {\bibfnamefont {I.~P.}\ \bibnamefont {McCulloch}}, \bibinfo {author}
  {\bibfnamefont {U.}~\bibnamefont {Schollw\"ock}}, \bibinfo {author}
  {\bibfnamefont {J.}~\bibnamefont {Eisert}}, \ and\ \bibinfo {author}
  {\bibfnamefont {I.}~\bibnamefont {Bloch}},\ }\bibfield  {title} {\enquote
  {\bibinfo {title} {Probing the relaxation towards equilibrium in an isolated
  strongly correlated {1D} {B}ose gas},}\ }\href@noop {} {\bibfield  {journal}
  {\bibinfo  {journal} {Nature Phys.}\ }\textbf {\bibinfo {volume} {8}},\
  \bibinfo {pages} {325} (\bibinfo {year} {2012})}\BibitemShut {NoStop}%
\bibitem [{\citenamefont {Gring}\ \emph {et~al.}(2012)\citenamefont {Gring},
  \citenamefont {Kuhnert}, \citenamefont {Langen}, \citenamefont {Kitagawa},
  \citenamefont {Rauer}, \citenamefont {Schreitl}, \citenamefont {Mazets},
  \citenamefont {Smith}, \citenamefont {Demler},\ and\ \citenamefont
  {Schmiedmayer}}]{gring_kuhnert_12}%
  \BibitemOpen
  \bibfield  {author} {\bibinfo {author} {\bibfnamefont {M.}~\bibnamefont
  {Gring}}, \bibinfo {author} {\bibfnamefont {M.}~\bibnamefont {Kuhnert}},
  \bibinfo {author} {\bibfnamefont {T.}~\bibnamefont {Langen}}, \bibinfo
  {author} {\bibfnamefont {T.}~\bibnamefont {Kitagawa}}, \bibinfo {author}
  {\bibfnamefont {B.}~\bibnamefont {Rauer}}, \bibinfo {author} {\bibfnamefont
  {M.}~\bibnamefont {Schreitl}}, \bibinfo {author} {\bibfnamefont
  {I.}~\bibnamefont {Mazets}}, \bibinfo {author} {\bibfnamefont {D.~Adu}\
  \bibnamefont {Smith}}, \bibinfo {author} {\bibfnamefont {E.}~\bibnamefont
  {Demler}}, \ and\ \bibinfo {author} {\bibfnamefont {J.}~\bibnamefont
  {Schmiedmayer}},\ }\bibfield  {title} {\enquote {\bibinfo {title} {Relaxation
  and prethermalization in an isolated quantum system},}\ }\href@noop {}
  {\bibfield  {journal} {\bibinfo  {journal} {Science}\ }\textbf {\bibinfo
  {volume} {337}},\ \bibinfo {pages} {1318--1322} (\bibinfo {year}
  {2012})}\BibitemShut {NoStop}%
\bibitem [{\citenamefont {Will}\ \emph {et~al.}(2015)\citenamefont {Will},
  \citenamefont {Iyer},\ and\ \citenamefont {Rigol}}]{will_iyer_15_97}%
  \BibitemOpen
  \bibfield  {author} {\bibinfo {author} {\bibfnamefont {S.}~\bibnamefont
  {Will}}, \bibinfo {author} {\bibfnamefont {D.}~\bibnamefont {Iyer}}, \ and\
  \bibinfo {author} {\bibfnamefont {M.}~\bibnamefont {Rigol}},\ }\bibfield
  {title} {\enquote {\bibinfo {title} {Observation of coherent quench dynamics
  in a metallic many-body state of fermionic atoms},}\ }\href@noop {}
  {\bibfield  {journal} {\bibinfo  {journal} {Nat. Commun.}\ }\textbf {\bibinfo
  {volume} {6}},\ \bibinfo {pages} {6009} (\bibinfo {year} {2015})}\BibitemShut
  {NoStop}%
\bibitem [{\citenamefont {Langen}\ \emph {et~al.}(2015)\citenamefont {Langen},
  \citenamefont {Erne}, \citenamefont {Geiger}, \citenamefont {Rauer},
  \citenamefont {Schweigler}, \citenamefont {Kuhnert}, \citenamefont
  {Rohringer}, \citenamefont {Mazets}, \citenamefont {Gasenzer},\ and\
  \citenamefont {Schmiedmayer}}]{langen_erne_15}%
  \BibitemOpen
  \bibfield  {author} {\bibinfo {author} {\bibfnamefont {T.}~\bibnamefont
  {Langen}}, \bibinfo {author} {\bibfnamefont {S.}~\bibnamefont {Erne}},
  \bibinfo {author} {\bibfnamefont {R.}~\bibnamefont {Geiger}}, \bibinfo
  {author} {\bibfnamefont {B.}~\bibnamefont {Rauer}}, \bibinfo {author}
  {\bibfnamefont {T.}~\bibnamefont {Schweigler}}, \bibinfo {author}
  {\bibfnamefont {M.}~\bibnamefont {Kuhnert}}, \bibinfo {author} {\bibfnamefont
  {W.}~\bibnamefont {Rohringer}}, \bibinfo {author} {\bibfnamefont {I.~E.}\
  \bibnamefont {Mazets}}, \bibinfo {author} {\bibfnamefont {T.}~\bibnamefont
  {Gasenzer}}, \ and\ \bibinfo {author} {\bibfnamefont {J.}~\bibnamefont
  {Schmiedmayer}},\ }\bibfield  {title} {\enquote {\bibinfo {title}
  {Experimental observation of a generalized {G}ibbs ensemble},}\ }\href@noop
  {} {\bibfield  {journal} {\bibinfo  {journal} {Science}\ }\textbf {\bibinfo
  {volume} {348}},\ \bibinfo {pages} {207} (\bibinfo {year}
  {2015})}\BibitemShut {NoStop}%
\bibitem [{\citenamefont {Schreiber}\ \emph {et~al.}(2015)\citenamefont
  {Schreiber}, \citenamefont {Hodgman}, \citenamefont {Bordia}, \citenamefont
  {L\"uschen}, \citenamefont {Fischer}, \citenamefont {Vosk}, \citenamefont
  {Altman}, \citenamefont {Schneider},\ and\ \citenamefont
  {Bloch}}]{schreiber_hodgman_15}%
  \BibitemOpen
  \bibfield  {author} {\bibinfo {author} {\bibfnamefont {M.}~\bibnamefont
  {Schreiber}}, \bibinfo {author} {\bibfnamefont {S.~S.}\ \bibnamefont
  {Hodgman}}, \bibinfo {author} {\bibfnamefont {P.}~\bibnamefont {Bordia}},
  \bibinfo {author} {\bibfnamefont {H.~P.}\ \bibnamefont {L\"uschen}}, \bibinfo
  {author} {\bibfnamefont {M.~H.}\ \bibnamefont {Fischer}}, \bibinfo {author}
  {\bibfnamefont {R.}~\bibnamefont {Vosk}}, \bibinfo {author} {\bibfnamefont
  {E.}~\bibnamefont {Altman}}, \bibinfo {author} {\bibfnamefont
  {U.}~\bibnamefont {Schneider}}, \ and\ \bibinfo {author} {\bibfnamefont
  {I.}~\bibnamefont {Bloch}},\ }\bibfield  {title} {\enquote {\bibinfo {title}
  {Observation of many-body localization of interacting fermions in a
  quasi-random optical lattice},}\ }\href@noop {} {\bibfield  {journal}
  {\bibinfo  {journal} {Science}\ }\textbf {\bibinfo {volume} {349}},\ \bibinfo
  {pages} {842} (\bibinfo {year} {2015})}\BibitemShut {NoStop}%
\bibitem [{\citenamefont {Rigol}\ \emph {et~al.}(2008)\citenamefont {Rigol},
  \citenamefont {Dunjko},\ and\ \citenamefont {Olshanii}}]{rigol_dunjko_08_34}%
  \BibitemOpen
  \bibfield  {author} {\bibinfo {author} {\bibfnamefont {M.}~\bibnamefont
  {Rigol}}, \bibinfo {author} {\bibfnamefont {V.}~\bibnamefont {Dunjko}}, \
  and\ \bibinfo {author} {\bibfnamefont {M.}~\bibnamefont {Olshanii}},\
  }\bibfield  {title} {\enquote {\bibinfo {title} {Thermalization and its
  mechanism for generic isolated quantum systems},}\ }\href@noop {} {\bibfield
  {journal} {\bibinfo  {journal} {Nature}\ }\textbf {\bibinfo {volume} {452}},\
  \bibinfo {pages} {854} (\bibinfo {year} {2008})}\BibitemShut {NoStop}%
\bibitem [{\citenamefont {Rigol}(2009{\natexlab{a}})}]{rigol_09_39}%
  \BibitemOpen
  \bibfield  {author} {\bibinfo {author} {\bibfnamefont {M.}~\bibnamefont
  {Rigol}},\ }\bibfield  {title} {\enquote {\bibinfo {title} {Breakdown of
  thermalization in finite one-dimensional systems},}\ }\href {\doibase
  10.1103/PhysRevLett.103.100403} {\bibfield  {journal} {\bibinfo  {journal}
  {Phys. Rev. Lett.}\ }\textbf {\bibinfo {volume} {103}},\ \bibinfo {pages}
  {100403} (\bibinfo {year} {2009}{\natexlab{a}})}\BibitemShut {NoStop}%
\bibitem [{\citenamefont {Rigol}(2009{\natexlab{b}})}]{rigol_09_43}%
  \BibitemOpen
  \bibfield  {author} {\bibinfo {author} {\bibfnamefont {M.}~\bibnamefont
  {Rigol}},\ }\bibfield  {title} {\enquote {\bibinfo {title} {Quantum quenches
  and thermalization in one-dimensional fermionic systems},}\ }\href {\doibase
  10.1103/PhysRevA.80.053607} {\bibfield  {journal} {\bibinfo  {journal} {Phys.
  Rev. A}\ }\textbf {\bibinfo {volume} {80}},\ \bibinfo {pages} {053607}
  (\bibinfo {year} {2009}{\natexlab{b}})}\BibitemShut {NoStop}%
\bibitem [{\citenamefont {Eckstein}\ \emph {et~al.}(2009)\citenamefont
  {Eckstein}, \citenamefont {Kollar},\ and\ \citenamefont
  {Werner}}]{eckstein_kollar_09}%
  \BibitemOpen
  \bibfield  {author} {\bibinfo {author} {\bibfnamefont {M.}~\bibnamefont
  {Eckstein}}, \bibinfo {author} {\bibfnamefont {M.}~\bibnamefont {Kollar}}, \
  and\ \bibinfo {author} {\bibfnamefont {P.}~\bibnamefont {Werner}},\
  }\bibfield  {title} {\enquote {\bibinfo {title} {Thermalization after an
  interaction quench in the {Hubbard} model},}\ }\href {\doibase
  10.1103/PhysRevLett.103.056403} {\bibfield  {journal} {\bibinfo  {journal}
  {Phys. Rev. Lett.}\ }\textbf {\bibinfo {volume} {103}},\ \bibinfo {pages}
  {056403} (\bibinfo {year} {2009})}\BibitemShut {NoStop}%
\bibitem [{\citenamefont {Ba\~nuls}\ \emph {et~al.}(2011)\citenamefont
  {Ba\~nuls}, \citenamefont {Cirac},\ and\ \citenamefont
  {Hastings}}]{banuls_cirac_11}%
  \BibitemOpen
  \bibfield  {author} {\bibinfo {author} {\bibfnamefont {M.~C.}\ \bibnamefont
  {Ba\~nuls}}, \bibinfo {author} {\bibfnamefont {J.~I.}\ \bibnamefont {Cirac}},
  \ and\ \bibinfo {author} {\bibfnamefont {M.~B.}\ \bibnamefont {Hastings}},\
  }\bibfield  {title} {\enquote {\bibinfo {title} {Strong and weak
  thermalization of infinite nonintegrable quantum systems},}\ }\href {\doibase
  10.1103/PhysRevLett.106.050405} {\bibfield  {journal} {\bibinfo  {journal}
  {Phys. Rev. Lett.}\ }\textbf {\bibinfo {volume} {106}},\ \bibinfo {pages}
  {050405} (\bibinfo {year} {2011})}\BibitemShut {NoStop}%
\bibitem [{\citenamefont {Khatami}\ \emph {et~al.}(2013)\citenamefont
  {Khatami}, \citenamefont {Pupillo}, \citenamefont {Srednicki},\ and\
  \citenamefont {Rigol}}]{khatami_pupillo_13_84}%
  \BibitemOpen
  \bibfield  {author} {\bibinfo {author} {\bibfnamefont {E.}~\bibnamefont
  {Khatami}}, \bibinfo {author} {\bibfnamefont {G.}~\bibnamefont {Pupillo}},
  \bibinfo {author} {\bibfnamefont {M.}~\bibnamefont {Srednicki}}, \ and\
  \bibinfo {author} {\bibfnamefont {M.}~\bibnamefont {Rigol}},\ }\bibfield
  {title} {\enquote {\bibinfo {title} {Fluctuation-dissipation theorem in an
  isolated system of quantum dipolar bosons after a quench},}\ }\href {\doibase
  10.1103/PhysRevLett.111.050403} {\bibfield  {journal} {\bibinfo  {journal}
  {Phys. Rev. Lett.}\ }\textbf {\bibinfo {volume} {111}},\ \bibinfo {pages}
  {050403} (\bibinfo {year} {2013})}\BibitemShut {NoStop}%
\bibitem [{\citenamefont {Sorg}\ \emph {et~al.}(2014)\citenamefont {Sorg},
  \citenamefont {Vidmar}, \citenamefont {Pollet},\ and\ \citenamefont
  {Heidrich-Meisner}}]{sorg_vidmar_14}%
  \BibitemOpen
  \bibfield  {author} {\bibinfo {author} {\bibfnamefont {S.}~\bibnamefont
  {Sorg}}, \bibinfo {author} {\bibfnamefont {L.}~\bibnamefont {Vidmar}},
  \bibinfo {author} {\bibfnamefont {L.}~\bibnamefont {Pollet}}, \ and\ \bibinfo
  {author} {\bibfnamefont {F.}~\bibnamefont {Heidrich-Meisner}},\ }\bibfield
  {title} {\enquote {\bibinfo {title} {Relaxation and thermalization in the
  one-dimensional {Bose-Hubbard} model: {A} case study for the interaction
  quantum quench from the atomic limit},}\ }\href {\doibase
  10.1103/PhysRevA.90.033606} {\bibfield  {journal} {\bibinfo  {journal} {Phys.
  Rev. A}\ }\textbf {\bibinfo {volume} {90}},\ \bibinfo {pages} {033606}
  (\bibinfo {year} {2014})}\BibitemShut {NoStop}%
\bibitem [{\citenamefont {Deutsch}(1991)}]{deutsch_91}%
  \BibitemOpen
  \bibfield  {author} {\bibinfo {author} {\bibfnamefont {J.~M.}\ \bibnamefont
  {Deutsch}},\ }\bibfield  {title} {\enquote {\bibinfo {title} {Quantum
  statistical mechanics in a closed system},}\ }\href {\doibase
  10.1103/PhysRevA.43.2046} {\bibfield  {journal} {\bibinfo  {journal} {Phys.
  Rev. A}\ }\textbf {\bibinfo {volume} {43}},\ \bibinfo {pages} {2046--2049}
  (\bibinfo {year} {1991})}\BibitemShut {NoStop}%
\bibitem [{\citenamefont {Srednicki}(1994)}]{srednicki_94}%
  \BibitemOpen
  \bibfield  {author} {\bibinfo {author} {\bibfnamefont {M.}~\bibnamefont
  {Srednicki}},\ }\bibfield  {title} {\enquote {\bibinfo {title} {Chaos and
  quantum thermalization},}\ }\href {\doibase 10.1103/PhysRevE.50.888}
  {\bibfield  {journal} {\bibinfo  {journal} {Phys. Rev. E}\ }\textbf {\bibinfo
  {volume} {50}},\ \bibinfo {pages} {888--901} (\bibinfo {year}
  {1994})}\BibitemShut {NoStop}%
\bibitem [{\citenamefont {Srednicki}(1999)}]{srednicki_99}%
  \BibitemOpen
  \bibfield  {author} {\bibinfo {author} {\bibfnamefont {M.}~\bibnamefont
  {Srednicki}},\ }\bibfield  {title} {\enquote {\bibinfo {title} {The approach
  to thermal equilibrium in quantized chaotic systems},}\ }\href@noop {}
  {\bibfield  {journal} {\bibinfo  {journal} {J. Phys. A}\ }\textbf {\bibinfo
  {volume} {32}},\ \bibinfo {pages} {1163} (\bibinfo {year}
  {1999})}\BibitemShut {NoStop}%
\bibitem [{\citenamefont {Rigol}(2014{\natexlab{a}})}]{rigol_14_90}%
  \BibitemOpen
  \bibfield  {author} {\bibinfo {author} {\bibfnamefont {M.}~\bibnamefont
  {Rigol}},\ }\bibfield  {title} {\enquote {\bibinfo {title} {Quantum quenches
  in the thermodynamic limit},}\ }\href {\doibase
  10.1103/PhysRevLett.112.170601} {\bibfield  {journal} {\bibinfo  {journal}
  {Phys. Rev. Lett.}\ }\textbf {\bibinfo {volume} {112}},\ \bibinfo {pages}
  {170601} (\bibinfo {year} {2014}{\natexlab{a}})}\BibitemShut {NoStop}%
\bibitem [{\citenamefont {Rigol}\ \emph {et~al.}(2007)\citenamefont {Rigol},
  \citenamefont {Dunjko}, \citenamefont {Yurovsky},\ and\ \citenamefont
  {Olshanii}}]{rigol_dunjko_07_27}%
  \BibitemOpen
  \bibfield  {author} {\bibinfo {author} {\bibfnamefont {M.}~\bibnamefont
  {Rigol}}, \bibinfo {author} {\bibfnamefont {V.}~\bibnamefont {Dunjko}},
  \bibinfo {author} {\bibfnamefont {V.}~\bibnamefont {Yurovsky}}, \ and\
  \bibinfo {author} {\bibfnamefont {M.}~\bibnamefont {Olshanii}},\ }\bibfield
  {title} {\enquote {\bibinfo {title} {Relaxation in a completely integrable
  many-body quantum system: {A}n \textit{Ab Initio} study of the dynamics of
  the highly excited states of {1D} lattice hard-core bosons},}\ }\href
  {\doibase 10.1103/PhysRevLett.98.050405} {\bibfield  {journal} {\bibinfo
  {journal} {Phys. Rev. Lett.}\ }\textbf {\bibinfo {volume} {98}},\ \bibinfo
  {pages} {050405} (\bibinfo {year} {2007})}\BibitemShut {NoStop}%
\bibitem [{\citenamefont {Cazalilla}(2006)}]{cazalilla_06}%
  \BibitemOpen
  \bibfield  {author} {\bibinfo {author} {\bibfnamefont {M.~A.}\ \bibnamefont
  {Cazalilla}},\ }\bibfield  {title} {\enquote {\bibinfo {title} {Effect of
  suddenly turning on interactions in the {Luttinger} model},}\ }\href
  {\doibase 10.1103/PhysRevLett.97.156403} {\bibfield  {journal} {\bibinfo
  {journal} {Phys. Rev. Lett.}\ }\textbf {\bibinfo {volume} {97}},\ \bibinfo
  {pages} {156403} (\bibinfo {year} {2006})}\BibitemShut {NoStop}%
\bibitem [{\citenamefont {Rigol}\ \emph
  {et~al.}(2006{\natexlab{a}})\citenamefont {Rigol}, \citenamefont
  {Muramatsu},\ and\ \citenamefont {Olshanii}}]{rigol_muramatsu_06_26}%
  \BibitemOpen
  \bibfield  {author} {\bibinfo {author} {\bibfnamefont {M.}~\bibnamefont
  {Rigol}}, \bibinfo {author} {\bibfnamefont {A.}~\bibnamefont {Muramatsu}}, \
  and\ \bibinfo {author} {\bibfnamefont {M.}~\bibnamefont {Olshanii}},\
  }\bibfield  {title} {\enquote {\bibinfo {title} {Hard-core bosons on optical
  superlattices: {D}ynamics and relaxation in the superfluid and insulating
  regimes},}\ }\href {\doibase 10.1103/PhysRevA.74.053616} {\bibfield
  {journal} {\bibinfo  {journal} {Phys. Rev. A}\ }\textbf {\bibinfo {volume}
  {74}},\ \bibinfo {pages} {053616} (\bibinfo {year}
  {2006}{\natexlab{a}})}\BibitemShut {NoStop}%
\bibitem [{\citenamefont {Kollar}\ and\ \citenamefont
  {Eckstein}(2008)}]{kollar_eckstein_08}%
  \BibitemOpen
  \bibfield  {author} {\bibinfo {author} {\bibfnamefont {M.}~\bibnamefont
  {Kollar}}\ and\ \bibinfo {author} {\bibfnamefont {M.}~\bibnamefont
  {Eckstein}},\ }\bibfield  {title} {\enquote {\bibinfo {title} {Relaxation of
  a one-dimensional mott insulator after an interaction quench},}\ }\href
  {\doibase 10.1103/PhysRevA.78.013626} {\bibfield  {journal} {\bibinfo
  {journal} {Phys. Rev. A}\ }\textbf {\bibinfo {volume} {78}},\ \bibinfo
  {pages} {013626} (\bibinfo {year} {2008})}\BibitemShut {NoStop}%
\bibitem [{\citenamefont {Barthel}\ and\ \citenamefont
  {Schollw\"ock}(2008)}]{barthel_schollwock_08}%
  \BibitemOpen
  \bibfield  {author} {\bibinfo {author} {\bibfnamefont {T.}~\bibnamefont
  {Barthel}}\ and\ \bibinfo {author} {\bibfnamefont {U.}~\bibnamefont
  {Schollw\"ock}},\ }\bibfield  {title} {\enquote {\bibinfo {title} {Dephasing
  and the steady state in quantum many-particle systems},}\ }\href {\doibase
  10.1103/PhysRevLett.100.100601} {\bibfield  {journal} {\bibinfo  {journal}
  {Phys. Rev. Lett.}\ }\textbf {\bibinfo {volume} {100}},\ \bibinfo {pages}
  {100601} (\bibinfo {year} {2008})}\BibitemShut {NoStop}%
\bibitem [{\citenamefont {Iucci}\ and\ \citenamefont
  {Cazalilla}(2009)}]{iucci_cazalilla_09}%
  \BibitemOpen
  \bibfield  {author} {\bibinfo {author} {\bibfnamefont {A.}~\bibnamefont
  {Iucci}}\ and\ \bibinfo {author} {\bibfnamefont {M.~A.}\ \bibnamefont
  {Cazalilla}},\ }\bibfield  {title} {\enquote {\bibinfo {title} {Quantum
  quench dynamics of the {Luttinger} model},}\ }\href {\doibase
  10.1103/PhysRevA.80.063619} {\bibfield  {journal} {\bibinfo  {journal} {Phys.
  Rev. A}\ }\textbf {\bibinfo {volume} {80}},\ \bibinfo {pages} {063619}
  (\bibinfo {year} {2009})}\BibitemShut {NoStop}%
\bibitem [{\citenamefont {Cassidy}\ \emph {et~al.}(2011)\citenamefont
  {Cassidy}, \citenamefont {Clark},\ and\ \citenamefont
  {Rigol}}]{cassidy_clark_11_55}%
  \BibitemOpen
  \bibfield  {author} {\bibinfo {author} {\bibfnamefont {A.~C.}\ \bibnamefont
  {Cassidy}}, \bibinfo {author} {\bibfnamefont {C.~W.}\ \bibnamefont {Clark}},
  \ and\ \bibinfo {author} {\bibfnamefont {M.}~\bibnamefont {Rigol}},\
  }\bibfield  {title} {\enquote {\bibinfo {title} {Generalized thermalization
  in an integrable lattice system},}\ }\href {\doibase
  10.1103/PhysRevLett.106.140405} {\bibfield  {journal} {\bibinfo  {journal}
  {Phys. Rev. Lett.}\ }\textbf {\bibinfo {volume} {106}},\ \bibinfo {pages}
  {140405} (\bibinfo {year} {2011})}\BibitemShut {NoStop}%
\bibitem [{\citenamefont {Calabrese}\ \emph {et~al.}(2011)\citenamefont
  {Calabrese}, \citenamefont {Essler},\ and\ \citenamefont
  {Fagotti}}]{calabrese_essler_11}%
  \BibitemOpen
  \bibfield  {author} {\bibinfo {author} {\bibfnamefont {P.}~\bibnamefont
  {Calabrese}}, \bibinfo {author} {\bibfnamefont {F.~H.~L.}\ \bibnamefont
  {Essler}}, \ and\ \bibinfo {author} {\bibfnamefont {M.}~\bibnamefont
  {Fagotti}},\ }\bibfield  {title} {\enquote {\bibinfo {title} {Quantum quench
  in the transverse-field {I}sing chain},}\ }\href {\doibase
  10.1103/PhysRevLett.106.227203} {\bibfield  {journal} {\bibinfo  {journal}
  {Phys. Rev. Lett.}\ }\textbf {\bibinfo {volume} {106}},\ \bibinfo {pages}
  {227203} (\bibinfo {year} {2011})}\BibitemShut {NoStop}%
\bibitem [{\citenamefont {Cazalilla}\ \emph {et~al.}(2012)\citenamefont
  {Cazalilla}, \citenamefont {Iucci},\ and\ \citenamefont
  {Chung}}]{cazalilla_iucci_12}%
  \BibitemOpen
  \bibfield  {author} {\bibinfo {author} {\bibfnamefont {M.~A.}\ \bibnamefont
  {Cazalilla}}, \bibinfo {author} {\bibfnamefont {A.}~\bibnamefont {Iucci}}, \
  and\ \bibinfo {author} {\bibfnamefont {M.-C.}\ \bibnamefont {Chung}},\
  }\bibfield  {title} {\enquote {\bibinfo {title} {Thermalization and quantum
  correlations in exactly solvable models},}\ }\href {\doibase
  10.1103/PhysRevE.85.011133} {\bibfield  {journal} {\bibinfo  {journal} {Phys.
  Rev. E}\ }\textbf {\bibinfo {volume} {85}},\ \bibinfo {pages} {011133}
  (\bibinfo {year} {2012})}\BibitemShut {NoStop}%
\bibitem [{\citenamefont {Caux}\ and\ \citenamefont
  {Konik}(2012)}]{caux_konik_12}%
  \BibitemOpen
  \bibfield  {author} {\bibinfo {author} {\bibfnamefont {J.-S.}\ \bibnamefont
  {Caux}}\ and\ \bibinfo {author} {\bibfnamefont {R.~M.}\ \bibnamefont
  {Konik}},\ }\bibfield  {title} {\enquote {\bibinfo {title} {Constructing the
  generalized {G}ibbs ensemble after a quantum quench},}\ }\href {\doibase
  10.1103/PhysRevLett.109.175301} {\bibfield  {journal} {\bibinfo  {journal}
  {Phys. Rev. Lett.}\ }\textbf {\bibinfo {volume} {109}},\ \bibinfo {pages}
  {175301} (\bibinfo {year} {2012})}\BibitemShut {NoStop}%
\bibitem [{\citenamefont {Gramsch}\ and\ \citenamefont
  {Rigol}(2012)}]{gramsch_rigol_12_77}%
  \BibitemOpen
  \bibfield  {author} {\bibinfo {author} {\bibfnamefont {C.}~\bibnamefont
  {Gramsch}}\ and\ \bibinfo {author} {\bibfnamefont {M.}~\bibnamefont
  {Rigol}},\ }\bibfield  {title} {\enquote {\bibinfo {title} {Quenches in a
  quasidisordered integrable lattice system: {D}ynamics and statistical
  description of observables after relaxation},}\ }\href {\doibase
  10.1103/PhysRevA.86.053615} {\bibfield  {journal} {\bibinfo  {journal} {Phys.
  Rev. A}\ }\textbf {\bibinfo {volume} {86}},\ \bibinfo {pages} {053615}
  (\bibinfo {year} {2012})}\BibitemShut {NoStop}%
\bibitem [{\citenamefont {Essler}\ \emph {et~al.}(2012)\citenamefont {Essler},
  \citenamefont {Evangelisti},\ and\ \citenamefont
  {Fagotti}}]{essler_evangelisti_12}%
  \BibitemOpen
  \bibfield  {author} {\bibinfo {author} {\bibfnamefont {F.~H.~L.}\
  \bibnamefont {Essler}}, \bibinfo {author} {\bibfnamefont {S.}~\bibnamefont
  {Evangelisti}}, \ and\ \bibinfo {author} {\bibfnamefont {M.}~\bibnamefont
  {Fagotti}},\ }\bibfield  {title} {\enquote {\bibinfo {title} {Dynamical
  correlations after a quantum quench},}\ }\href {\doibase
  10.1103/PhysRevLett.109.247206} {\bibfield  {journal} {\bibinfo  {journal}
  {Phys. Rev. Lett.}\ }\textbf {\bibinfo {volume} {109}},\ \bibinfo {pages}
  {247206} (\bibinfo {year} {2012})}\BibitemShut {NoStop}%
\bibitem [{\citenamefont {Collura}\ \emph {et~al.}(2013)\citenamefont
  {Collura}, \citenamefont {Sotiriadis},\ and\ \citenamefont
  {Calabrese}}]{collura_sotiriadis_13a}%
  \BibitemOpen
  \bibfield  {author} {\bibinfo {author} {\bibfnamefont {M.}~\bibnamefont
  {Collura}}, \bibinfo {author} {\bibfnamefont {S.}~\bibnamefont {Sotiriadis}},
  \ and\ \bibinfo {author} {\bibfnamefont {P.}~\bibnamefont {Calabrese}},\
  }\bibfield  {title} {\enquote {\bibinfo {title} {Equilibration of a
  {Tonks-Girardeau} gas following a trap release},}\ }\href {\doibase
  10.1103/PhysRevLett.110.245301} {\bibfield  {journal} {\bibinfo  {journal}
  {Phys. Rev. Lett.}\ }\textbf {\bibinfo {volume} {110}},\ \bibinfo {pages}
  {245301} (\bibinfo {year} {2013})}\BibitemShut {NoStop}%
\bibitem [{\citenamefont {Caux}\ and\ \citenamefont
  {Essler}(2013)}]{caux_essler_13}%
  \BibitemOpen
  \bibfield  {author} {\bibinfo {author} {\bibfnamefont {J.-S.}\ \bibnamefont
  {Caux}}\ and\ \bibinfo {author} {\bibfnamefont {F.~H.~L.}\ \bibnamefont
  {Essler}},\ }\bibfield  {title} {\enquote {\bibinfo {title} {Time evolution
  of local observables after quenching to an integrable model},}\ }\href
  {\doibase 10.1103/PhysRevLett.110.257203} {\bibfield  {journal} {\bibinfo
  {journal} {Phys. Rev. Lett.}\ }\textbf {\bibinfo {volume} {110}},\ \bibinfo
  {pages} {257203} (\bibinfo {year} {2013})}\BibitemShut {NoStop}%
\bibitem [{\citenamefont {Mussardo}(2013)}]{mussardo_13}%
  \BibitemOpen
  \bibfield  {author} {\bibinfo {author} {\bibfnamefont {G.}~\bibnamefont
  {Mussardo}},\ }\bibfield  {title} {\enquote {\bibinfo {title} {Infinite-time
  average of local fields in an integrable quantum field theory after a quantum
  quench},}\ }\href {\doibase 10.1103/PhysRevLett.111.100401} {\bibfield
  {journal} {\bibinfo  {journal} {Phys. Rev. Lett.}\ }\textbf {\bibinfo
  {volume} {111}},\ \bibinfo {pages} {100401} (\bibinfo {year}
  {2013})}\BibitemShut {NoStop}%
\bibitem [{\citenamefont {Fagotti}\ and\ \citenamefont
  {Essler}(2013)}]{fagotti_essler_13a}%
  \BibitemOpen
  \bibfield  {author} {\bibinfo {author} {\bibfnamefont {M.}~\bibnamefont
  {Fagotti}}\ and\ \bibinfo {author} {\bibfnamefont {F.~H.~L.}\ \bibnamefont
  {Essler}},\ }\bibfield  {title} {\enquote {\bibinfo {title} {Reduced density
  matrix after a quantum quench},}\ }\href {\doibase
  10.1103/PhysRevB.87.245107} {\bibfield  {journal} {\bibinfo  {journal} {Phys.
  Rev. B}\ }\textbf {\bibinfo {volume} {87}},\ \bibinfo {pages} {245107}
  (\bibinfo {year} {2013})}\BibitemShut {NoStop}%
\bibitem [{\citenamefont {Wright}\ \emph {et~al.}(2014)\citenamefont {Wright},
  \citenamefont {Rigol}, \citenamefont {Davis},\ and\ \citenamefont
  {Kheruntsyan}}]{wright_rigol_14_92}%
  \BibitemOpen
  \bibfield  {author} {\bibinfo {author} {\bibfnamefont {T.~M.}\ \bibnamefont
  {Wright}}, \bibinfo {author} {\bibfnamefont {M.}~\bibnamefont {Rigol}},
  \bibinfo {author} {\bibfnamefont {M.~J.}\ \bibnamefont {Davis}}, \ and\
  \bibinfo {author} {\bibfnamefont {K.~V.}\ \bibnamefont {Kheruntsyan}},\
  }\bibfield  {title} {\enquote {\bibinfo {title} {Nonequilibrium dynamics of
  one-dimensional hard-core anyons following a quench: {C}omplete relaxation of
  one-body observables},}\ }\href {\doibase 10.1103/PhysRevLett.113.050601}
  {\bibfield  {journal} {\bibinfo  {journal} {Phys. Rev. Lett.}\ }\textbf
  {\bibinfo {volume} {113}},\ \bibinfo {pages} {050601} (\bibinfo {year}
  {2014})}\BibitemShut {NoStop}%
\bibitem [{\citenamefont {Pozsgay}(2014{\natexlab{a}})}]{pozsgay_14a}%
  \BibitemOpen
  \bibfield  {author} {\bibinfo {author} {\bibfnamefont {B.}~\bibnamefont
  {Pozsgay}},\ }\bibfield  {title} {\enquote {\bibinfo {title} {Failure of the
  generalized eigenstate thermalization hypothesis in integrable models with
  multiple particle species},}\ }\href@noop {} {\bibfield  {journal} {\bibinfo
  {journal} {J. Stat. Mech.}\ ,\ \bibinfo {pages} {P09026}} (\bibinfo {year}
  {2014}{\natexlab{a}})}\BibitemShut {NoStop}%
\bibitem [{\citenamefont {Pozsgay}(2014{\natexlab{b}})}]{pozsgay_14b}%
  \BibitemOpen
  \bibfield  {author} {\bibinfo {author} {\bibfnamefont {B.}~\bibnamefont
  {Pozsgay}},\ }\bibfield  {title} {\enquote {\bibinfo {title} {Quantum
  quenches and generalized {G}ibbs ensemble in a {Bethe Ansatz} solvable
  lattice model of interacting bosons},}\ }\href@noop {} {\bibfield  {journal}
  {\bibinfo  {journal} {J. Stat. Mech.}\ ,\ \bibinfo {pages} {P10045}}
  (\bibinfo {year} {2014}{\natexlab{b}})}\BibitemShut {NoStop}%
\bibitem [{\citenamefont {Cardy}()}]{cardy_16}%
  \BibitemOpen
  \bibfield  {author} {\bibinfo {author} {\bibfnamefont {J.}~\bibnamefont
  {Cardy}},\ }\href@noop {} {\enquote {\bibinfo {title} {Quantum quenches to a
  critical point in one dimension: some further results},}\ }\bibinfo {note}
  {{a}rXiv:1507.07266}\BibitemShut {NoStop}%
\bibitem [{\citenamefont {Tasaki}(1998)}]{tasaki_98}%
  \BibitemOpen
  \bibfield  {author} {\bibinfo {author} {\bibfnamefont {H.}~\bibnamefont
  {Tasaki}},\ }\bibfield  {title} {\enquote {\bibinfo {title} {From quantum
  dynamics to the canonical distribution: General picture and a rigorous
  example},}\ }\href {\doibase 10.1103/PhysRevLett.80.1373} {\bibfield
  {journal} {\bibinfo  {journal} {Phys. Rev. Lett.}\ }\textbf {\bibinfo
  {volume} {80}},\ \bibinfo {pages} {1373--1376} (\bibinfo {year}
  {1998})}\BibitemShut {NoStop}%
\bibitem [{\citenamefont {Goldstein}\ \emph {et~al.}(2006)\citenamefont
  {Goldstein}, \citenamefont {Lebowitz}, \citenamefont {Tumulka},\ and\
  \citenamefont {Zangh\`\i{}}}]{goldstein_lebowitz_06}%
  \BibitemOpen
  \bibfield  {author} {\bibinfo {author} {\bibfnamefont {S.}~\bibnamefont
  {Goldstein}}, \bibinfo {author} {\bibfnamefont {J.~L.}\ \bibnamefont
  {Lebowitz}}, \bibinfo {author} {\bibfnamefont {R.}~\bibnamefont {Tumulka}}, \
  and\ \bibinfo {author} {\bibfnamefont {N.}~\bibnamefont {Zangh\`\i{}}},\
  }\bibfield  {title} {\enquote {\bibinfo {title} {Canonical typicality},}\
  }\href {\doibase 10.1103/PhysRevLett.96.050403} {\bibfield  {journal}
  {\bibinfo  {journal} {Phys. Rev. Lett.}\ }\textbf {\bibinfo {volume} {96}},\
  \bibinfo {pages} {050403} (\bibinfo {year} {2006})}\BibitemShut {NoStop}%
\bibitem [{\citenamefont {Popescu}\ \emph {et~al.}(2006)\citenamefont
  {Popescu}, \citenamefont {Short},\ and\ \citenamefont {Winter}}]{popescu_06}%
  \BibitemOpen
  \bibfield  {author} {\bibinfo {author} {\bibfnamefont {S.}~\bibnamefont
  {Popescu}}, \bibinfo {author} {\bibfnamefont {A.~J.}\ \bibnamefont {Short}},
  \ and\ \bibinfo {author} {\bibfnamefont {A.}~\bibnamefont {Winter}},\
  }\bibfield  {title} {\enquote {\bibinfo {title} {Entanglement and the
  foundations of statistical mechanics},}\ }\href@noop {} {\bibfield  {journal}
  {\bibinfo  {journal} {Nature Phys.}\ }\textbf {\bibinfo {volume} {2}},\
  \bibinfo {pages} {754} (\bibinfo {year} {2006})}\BibitemShut {NoStop}%
\bibitem [{\citenamefont {Eisert}\ \emph {et~al.}(2015)\citenamefont {Eisert},
  \citenamefont {Friesdorf},\ and\ \citenamefont
  {Gogolin}}]{eisert_friesdorf_review_15}%
  \BibitemOpen
  \bibfield  {author} {\bibinfo {author} {\bibfnamefont {J.}~\bibnamefont
  {Eisert}}, \bibinfo {author} {\bibfnamefont {M.}~\bibnamefont {Friesdorf}}, \
  and\ \bibinfo {author} {\bibfnamefont {C.}~\bibnamefont {Gogolin}},\
  }\bibfield  {title} {\enquote {\bibinfo {title} {Quantum many-body systems
  out of equilibrium},}\ }\href@noop {} {\bibfield  {journal} {\bibinfo
  {journal} {Nature Physics}\ }\textbf {\bibinfo {volume} {11}},\ \bibinfo
  {pages} {124} (\bibinfo {year} {2015})}\BibitemShut {NoStop}%
\bibitem [{\citenamefont {Wouters}\ \emph {et~al.}(2014)\citenamefont
  {Wouters}, \citenamefont {De~Nardis}, \citenamefont {Brockmann},
  \citenamefont {Fioretto}, \citenamefont {Rigol},\ and\ \citenamefont
  {Caux}}]{wouters_denardis_14_93}%
  \BibitemOpen
  \bibfield  {author} {\bibinfo {author} {\bibfnamefont {B.}~\bibnamefont
  {Wouters}}, \bibinfo {author} {\bibfnamefont {J.}~\bibnamefont {De~Nardis}},
  \bibinfo {author} {\bibfnamefont {M.}~\bibnamefont {Brockmann}}, \bibinfo
  {author} {\bibfnamefont {D.}~\bibnamefont {Fioretto}}, \bibinfo {author}
  {\bibfnamefont {M.}~\bibnamefont {Rigol}}, \ and\ \bibinfo {author}
  {\bibfnamefont {J.-S.}\ \bibnamefont {Caux}},\ }\bibfield  {title} {\enquote
  {\bibinfo {title} {Quenching the anisotropic {H}eisenberg chain: {E}xact
  solution and generalized {G}ibbs ensemble predictions},}\ }\href {\doibase
  10.1103/PhysRevLett.113.117202} {\bibfield  {journal} {\bibinfo  {journal}
  {Phys. Rev. Lett.}\ }\textbf {\bibinfo {volume} {113}},\ \bibinfo {pages}
  {117202} (\bibinfo {year} {2014})}\BibitemShut {NoStop}%
\bibitem [{\citenamefont {Rigol}(2014{\natexlab{b}})}]{rigol_14_95}%
  \BibitemOpen
  \bibfield  {author} {\bibinfo {author} {\bibfnamefont {M.}~\bibnamefont
  {Rigol}},\ }\bibfield  {title} {\enquote {\bibinfo {title} {Quantum quenches
  in the thermodynamic limit. {II. I}nitial ground states},}\ }\href {\doibase
  10.1103/PhysRevE.90.031301} {\bibfield  {journal} {\bibinfo  {journal} {Phys.
  Rev. E}\ }\textbf {\bibinfo {volume} {90}},\ \bibinfo {pages} {031301(R)}
  (\bibinfo {year} {2014}{\natexlab{b}})}\BibitemShut {NoStop}%
\bibitem [{\citenamefont {Rigol}\ \emph
  {et~al.}(2006{\natexlab{b}})\citenamefont {Rigol}, \citenamefont {Bryant},\
  and\ \citenamefont {Singh}}]{rigol_bryant_06_25}%
  \BibitemOpen
  \bibfield  {author} {\bibinfo {author} {\bibfnamefont {M.}~\bibnamefont
  {Rigol}}, \bibinfo {author} {\bibfnamefont {T.}~\bibnamefont {Bryant}}, \
  and\ \bibinfo {author} {\bibfnamefont {R.~R.~P.}\ \bibnamefont {Singh}},\
  }\bibfield  {title} {\enquote {\bibinfo {title} {Numerical linked-cluster
  approach to quantum lattice models},}\ }\href {\doibase
  10.1103/PhysRevLett.97.187202} {\bibfield  {journal} {\bibinfo  {journal}
  {Phys. Rev. Lett.}\ }\textbf {\bibinfo {volume} {97}},\ \bibinfo {pages}
  {187202} (\bibinfo {year} {2006}{\natexlab{b}})}\BibitemShut {NoStop}%
\bibitem [{\citenamefont {Cazalilla}\ \emph {et~al.}(2011)\citenamefont
  {Cazalilla}, \citenamefont {Citro}, \citenamefont {Giamarchi}, \citenamefont
  {Orignac},\ and\ \citenamefont {Rigol}}]{cazalilla_citro_review_11_63}%
  \BibitemOpen
  \bibfield  {author} {\bibinfo {author} {\bibfnamefont {M.~A.}\ \bibnamefont
  {Cazalilla}}, \bibinfo {author} {\bibfnamefont {R.}~\bibnamefont {Citro}},
  \bibinfo {author} {\bibfnamefont {T.}~\bibnamefont {Giamarchi}}, \bibinfo
  {author} {\bibfnamefont {E.}~\bibnamefont {Orignac}}, \ and\ \bibinfo
  {author} {\bibfnamefont {M.}~\bibnamefont {Rigol}},\ }\bibfield  {title}
  {\enquote {\bibinfo {title} {One dimensional bosons: {F}rom condensed matter
  systems to ultracold gases},}\ }\href {\doibase 10.1103/RevModPhys.83.1405}
  {\bibfield  {journal} {\bibinfo  {journal} {Rev. Mod. Phys.}\ }\textbf
  {\bibinfo {volume} {83}},\ \bibinfo {pages} {1405--1466} (\bibinfo {year}
  {2011})}\BibitemShut {NoStop}%
\bibitem [{\citenamefont {Pozsgay}(2013)}]{pozsgay_13}%
  \BibitemOpen
  \bibfield  {author} {\bibinfo {author} {\bibfnamefont {B.}~\bibnamefont
  {Pozsgay}},\ }\bibfield  {title} {\enquote {\bibinfo {title} {The generalized
  {G}ibbs ensemble for {H}eisenberg spin chains},}\ }\href@noop {} {\bibfield
  {journal} {\bibinfo  {journal} {J. Stat. Mech.}\ }\textbf {\bibinfo {volume}
  {2013}},\ \bibinfo {pages} {P07003} (\bibinfo {year} {2013})}\BibitemShut
  {NoStop}%
\bibitem [{\citenamefont {Fagotti}\ \emph {et~al.}(2014)\citenamefont
  {Fagotti}, \citenamefont {Collura}, \citenamefont {Essler},\ and\
  \citenamefont {Calabrese}}]{fagotti_collura_14}%
  \BibitemOpen
  \bibfield  {author} {\bibinfo {author} {\bibfnamefont {M.}~\bibnamefont
  {Fagotti}}, \bibinfo {author} {\bibfnamefont {M.}~\bibnamefont {Collura}},
  \bibinfo {author} {\bibfnamefont {F.~H.~L.}\ \bibnamefont {Essler}}, \ and\
  \bibinfo {author} {\bibfnamefont {P.}~\bibnamefont {Calabrese}},\ }\bibfield
  {title} {\enquote {\bibinfo {title} {Relaxation after quantum quenches in the
  spin-$\frac{1}{2}$ {H}eisenberg {XXZ} chain},}\ }\href {\doibase
  10.1103/PhysRevB.89.125101} {\bibfield  {journal} {\bibinfo  {journal} {Phys.
  Rev. B}\ }\textbf {\bibinfo {volume} {89}},\ \bibinfo {pages} {125101}
  (\bibinfo {year} {2014})}\BibitemShut {NoStop}%
\bibitem [{\citenamefont {Pozsgay}\ \emph {et~al.}(2014)\citenamefont
  {Pozsgay}, \citenamefont {Mesty\'an}, \citenamefont {Werner}, \citenamefont
  {Kormos}, \citenamefont {Zar\'and},\ and\ \citenamefont
  {Tak\'acs}}]{pozsgay_mestyan14}%
  \BibitemOpen
  \bibfield  {author} {\bibinfo {author} {\bibfnamefont {B.}~\bibnamefont
  {Pozsgay}}, \bibinfo {author} {\bibfnamefont {M.}~\bibnamefont {Mesty\'an}},
  \bibinfo {author} {\bibfnamefont {M.~A.}\ \bibnamefont {Werner}}, \bibinfo
  {author} {\bibfnamefont {M.}~\bibnamefont {Kormos}}, \bibinfo {author}
  {\bibfnamefont {G.}~\bibnamefont {Zar\'and}}, \ and\ \bibinfo {author}
  {\bibfnamefont {G.}~\bibnamefont {Tak\'acs}},\ }\bibfield  {title} {\enquote
  {\bibinfo {title} {Correlations after quantum quenches in the {XXZ} spin
  chain: {F}ailure of the generalized {G}ibbs ensemble},}\ }\href {\doibase
  10.1103/PhysRevLett.113.117203} {\bibfield  {journal} {\bibinfo  {journal}
  {Phys. Rev. Lett.}\ }\textbf {\bibinfo {volume} {113}},\ \bibinfo {pages}
  {117203} (\bibinfo {year} {2014})}\BibitemShut {NoStop}%
\bibitem [{\citenamefont {Mierzejewski}\ \emph {et~al.}(2014)\citenamefont
  {Mierzejewski}, \citenamefont {Prelov\ifmmode~\check{s}\else \v{s}\fi{}ek},\
  and\ \citenamefont {Prosen}}]{mierzejewski_prelovssek_14}%
  \BibitemOpen
  \bibfield  {author} {\bibinfo {author} {\bibfnamefont {M.}~\bibnamefont
  {Mierzejewski}}, \bibinfo {author} {\bibfnamefont {P.}~\bibnamefont
  {Prelov\ifmmode~\check{s}\else \v{s}\fi{}ek}}, \ and\ \bibinfo {author}
  {\bibfnamefont {T.}~\bibnamefont {Prosen}},\ }\bibfield  {title} {\enquote
  {\bibinfo {title} {Breakdown of the generalized {G}ibbs ensemble for
  current-generating quenches},}\ }\href {\doibase
  10.1103/PhysRevLett.113.020602} {\bibfield  {journal} {\bibinfo  {journal}
  {Phys. Rev. Lett.}\ }\textbf {\bibinfo {volume} {113}},\ \bibinfo {pages}
  {020602} (\bibinfo {year} {2014})}\BibitemShut {NoStop}%
\bibitem [{\citenamefont {Goldstein}\ and\ \citenamefont
  {Andrei}(2014)}]{goldstein_andrei_14}%
  \BibitemOpen
  \bibfield  {author} {\bibinfo {author} {\bibfnamefont {G.}~\bibnamefont
  {Goldstein}}\ and\ \bibinfo {author} {\bibfnamefont {N.}~\bibnamefont
  {Andrei}},\ }\bibfield  {title} {\enquote {\bibinfo {title} {Failure of the
  local generalized {G}ibbs ensemble for integrable models with bound
  states},}\ }\href {\doibase 10.1103/PhysRevA.90.043625} {\bibfield  {journal}
  {\bibinfo  {journal} {Phys. Rev. A}\ }\textbf {\bibinfo {volume} {90}},\
  \bibinfo {pages} {043625} (\bibinfo {year} {2014})}\BibitemShut {NoStop}%
\bibitem [{\citenamefont {Mierzejewski}\ \emph {et~al.}(2015)\citenamefont
  {Mierzejewski}, \citenamefont {Prosen},\ and\ \citenamefont
  {Prelov\ifmmode~\check{s}\else \v{s}\fi{}ek}}]{mierzejewski_prosen_15}%
  \BibitemOpen
  \bibfield  {author} {\bibinfo {author} {\bibfnamefont {M.}~\bibnamefont
  {Mierzejewski}}, \bibinfo {author} {\bibfnamefont {T.}~\bibnamefont
  {Prosen}}, \ and\ \bibinfo {author} {\bibfnamefont {P.}~\bibnamefont
  {Prelov\ifmmode~\check{s}\else \v{s}\fi{}ek}},\ }\bibfield  {title} {\enquote
  {\bibinfo {title} {Approximate conservation laws in perturbed integrable
  lattice models},}\ }\href {\doibase 10.1103/PhysRevB.92.195121} {\bibfield
  {journal} {\bibinfo  {journal} {Phys. Rev. B}\ }\textbf {\bibinfo {volume}
  {92}},\ \bibinfo {pages} {195121} (\bibinfo {year} {2015})}\BibitemShut
  {NoStop}%
\bibitem [{\citenamefont {Ilievski}\ \emph
  {et~al.}(2015{\natexlab{a}})\citenamefont {Ilievski}, \citenamefont
  {Medenjak},\ and\ \citenamefont {Prosen}}]{ilievski_medenjak_15}%
  \BibitemOpen
  \bibfield  {author} {\bibinfo {author} {\bibfnamefont {E.}~\bibnamefont
  {Ilievski}}, \bibinfo {author} {\bibfnamefont {M.}~\bibnamefont {Medenjak}},
  \ and\ \bibinfo {author} {\bibfnamefont {T.}~\bibnamefont {Prosen}},\
  }\bibfield  {title} {\enquote {\bibinfo {title} {Quasilocal conserved
  operators in the isotropic {H}eisenberg spin-$1/2$ chain},}\ }\href {\doibase
  10.1103/PhysRevLett.115.120601} {\bibfield  {journal} {\bibinfo  {journal}
  {Phys. Rev. Lett.}\ }\textbf {\bibinfo {volume} {115}},\ \bibinfo {pages}
  {120601} (\bibinfo {year} {2015}{\natexlab{a}})}\BibitemShut {NoStop}%
\bibitem [{\citenamefont {Ilievski}\ \emph
  {et~al.}(2015{\natexlab{b}})\citenamefont {Ilievski}, \citenamefont
  {De~Nardis}, \citenamefont {Wouters}, \citenamefont {Caux}, \citenamefont
  {Essler},\ and\ \citenamefont {Prosen}}]{ilievski_denardis_15}%
  \BibitemOpen
  \bibfield  {author} {\bibinfo {author} {\bibfnamefont {E.}~\bibnamefont
  {Ilievski}}, \bibinfo {author} {\bibfnamefont {J.}~\bibnamefont {De~Nardis}},
  \bibinfo {author} {\bibfnamefont {B.}~\bibnamefont {Wouters}}, \bibinfo
  {author} {\bibfnamefont {J.-S.}\ \bibnamefont {Caux}}, \bibinfo {author}
  {\bibfnamefont {F.~H.~L.}\ \bibnamefont {Essler}}, \ and\ \bibinfo {author}
  {\bibfnamefont {T.}~\bibnamefont {Prosen}},\ }\bibfield  {title} {\enquote
  {\bibinfo {title} {Complete generalized {G}ibbs ensembles in an interacting
  theory},}\ }\href {\doibase 10.1103/PhysRevLett.115.157201} {\bibfield
  {journal} {\bibinfo  {journal} {Phys. Rev. Lett.}\ }\textbf {\bibinfo
  {volume} {115}},\ \bibinfo {pages} {157201} (\bibinfo {year}
  {2015}{\natexlab{b}})}\BibitemShut {NoStop}%
\bibitem [{\citenamefont {Santos}\ \emph
  {et~al.}(2012{\natexlab{a}})\citenamefont {Santos}, \citenamefont
  {Borgonovi},\ and\ \citenamefont {Izrailev}}]{santos_borgonovi_12a}%
  \BibitemOpen
  \bibfield  {author} {\bibinfo {author} {\bibfnamefont {L.~F.}\ \bibnamefont
  {Santos}}, \bibinfo {author} {\bibfnamefont {F.}~\bibnamefont {Borgonovi}}, \
  and\ \bibinfo {author} {\bibfnamefont {F.~M.}\ \bibnamefont {Izrailev}},\
  }\bibfield  {title} {\enquote {\bibinfo {title} {Chaos and statistical
  relaxation in quantum systems of interacting particles},}\ }\href {\doibase
  10.1103/PhysRevLett.108.094102} {\bibfield  {journal} {\bibinfo  {journal}
  {Phys. Rev. Lett.}\ }\textbf {\bibinfo {volume} {108}},\ \bibinfo {pages}
  {094102} (\bibinfo {year} {2012}{\natexlab{a}})}\BibitemShut {NoStop}%
\bibitem [{\citenamefont {Santos}\ \emph
  {et~al.}(2012{\natexlab{b}})\citenamefont {Santos}, \citenamefont
  {Borgonovi},\ and\ \citenamefont {Izrailev}}]{santos_borgonovi_12b}%
  \BibitemOpen
  \bibfield  {author} {\bibinfo {author} {\bibfnamefont {L.~F.}\ \bibnamefont
  {Santos}}, \bibinfo {author} {\bibfnamefont {F.}~\bibnamefont {Borgonovi}}, \
  and\ \bibinfo {author} {\bibfnamefont {F.~M.}\ \bibnamefont {Izrailev}},\
  }\bibfield  {title} {\enquote {\bibinfo {title} {Onset of chaos and
  relaxation in isolated systems of interacting spins: Energy shell
  approach},}\ }\href {\doibase 10.1103/PhysRevE.85.036209} {\bibfield
  {journal} {\bibinfo  {journal} {Phys. Rev. E}\ }\textbf {\bibinfo {volume}
  {85}},\ \bibinfo {pages} {036209} (\bibinfo {year}
  {2012}{\natexlab{b}})}\BibitemShut {NoStop}%
\bibitem [{\citenamefont {He}\ and\ \citenamefont
  {Rigol}(2013)}]{he_rigol_13_81}%
  \BibitemOpen
  \bibfield  {author} {\bibinfo {author} {\bibfnamefont {K.}~\bibnamefont
  {He}}\ and\ \bibinfo {author} {\bibfnamefont {M.}~\bibnamefont {Rigol}},\
  }\bibfield  {title} {\enquote {\bibinfo {title} {Initial-state dependence of
  the quench dynamics in integrable quantum systems. {III. C}haotic states},}\
  }\href {\doibase 10.1103/PhysRevA.87.043615} {\bibfield  {journal} {\bibinfo
  {journal} {Phys. Rev. A}\ }\textbf {\bibinfo {volume} {87}},\ \bibinfo
  {pages} {043615} (\bibinfo {year} {2013})}\BibitemShut {NoStop}%
\bibitem [{\citenamefont {Polkovnikov}(2011)}]{polkovnikov_11}%
  \BibitemOpen
  \bibfield  {author} {\bibinfo {author} {\bibfnamefont {A.}~\bibnamefont
  {Polkovnikov}},\ }\bibfield  {title} {\enquote {\bibinfo {title} {Microscopic
  diagonal entropy and its connection to basic thermodynamic relations},}\
  }\href {\doibase DOI: 10.1016/j.aop.2010.08.004} {\bibfield  {journal}
  {\bibinfo  {journal} {Ann. Phys.}\ }\textbf {\bibinfo {volume} {326}},\
  \bibinfo {pages} {486 -- 499} (\bibinfo {year} {2011})}\BibitemShut {NoStop}%
\bibitem [{\citenamefont {Santos}\ \emph {et~al.}(2011)\citenamefont {Santos},
  \citenamefont {Polkovnikov},\ and\ \citenamefont
  {Rigol}}]{santos_polkovnikov_11_57}%
  \BibitemOpen
  \bibfield  {author} {\bibinfo {author} {\bibfnamefont {L.~F.}\ \bibnamefont
  {Santos}}, \bibinfo {author} {\bibfnamefont {A.}~\bibnamefont {Polkovnikov}},
  \ and\ \bibinfo {author} {\bibfnamefont {M.}~\bibnamefont {Rigol}},\
  }\bibfield  {title} {\enquote {\bibinfo {title} {Entropy of isolated quantum
  systems after a quench},}\ }\href {\doibase 10.1103/PhysRevLett.107.040601}
  {\bibfield  {journal} {\bibinfo  {journal} {Phys. Rev. Lett.}\ }\textbf
  {\bibinfo {volume} {107}},\ \bibinfo {pages} {040601} (\bibinfo {year}
  {2011})}\BibitemShut {NoStop}%
\bibitem [{\citenamefont {He}\ and\ \citenamefont
  {Rigol}(2012)}]{he_rigol_12_71}%
  \BibitemOpen
  \bibfield  {author} {\bibinfo {author} {\bibfnamefont {K.}~\bibnamefont
  {He}}\ and\ \bibinfo {author} {\bibfnamefont {M.}~\bibnamefont {Rigol}},\
  }\bibfield  {title} {\enquote {\bibinfo {title} {Initial-state dependence of
  the quench dynamics in integrable quantum systems. {II. T}hermal states},}\
  }\href {\doibase 10.1103/PhysRevA.85.063609} {\bibfield  {journal} {\bibinfo
  {journal} {Phys. Rev. A}\ }\textbf {\bibinfo {volume} {85}},\ \bibinfo
  {pages} {063609} (\bibinfo {year} {2012})}\BibitemShut {NoStop}%
\bibitem [{\citenamefont {Rigol}\ and\ \citenamefont
  {Srednicki}(2012)}]{rigol_srednicki_12_69}%
  \BibitemOpen
  \bibfield  {author} {\bibinfo {author} {\bibfnamefont {M.}~\bibnamefont
  {Rigol}}\ and\ \bibinfo {author} {\bibfnamefont {M.}~\bibnamefont
  {Srednicki}},\ }\bibfield  {title} {\enquote {\bibinfo {title} {Alternatives
  to eigenstate thermalization},}\ }\href {\doibase
  10.1103/PhysRevLett.108.110601} {\bibfield  {journal} {\bibinfo  {journal}
  {Phys. Rev. Lett.}\ }\textbf {\bibinfo {volume} {108}},\ \bibinfo {pages}
  {110601} (\bibinfo {year} {2012})}\BibitemShut {NoStop}%
\bibitem [{\citenamefont {Bloch}\ \emph {et~al.}(2008)\citenamefont {Bloch},
  \citenamefont {Dalibard},\ and\ \citenamefont
  {Zwerger}}]{bloch_dalibard_review_08}%
  \BibitemOpen
  \bibfield  {author} {\bibinfo {author} {\bibfnamefont {I.}~\bibnamefont
  {Bloch}}, \bibinfo {author} {\bibfnamefont {J.}~\bibnamefont {Dalibard}}, \
  and\ \bibinfo {author} {\bibfnamefont {W.}~\bibnamefont {Zwerger}},\
  }\bibfield  {title} {\enquote {\bibinfo {title} {Many-body physics with
  ultracold gases},}\ }\href {\doibase 10.1103/RevModPhys.80.885} {\bibfield
  {journal} {\bibinfo  {journal} {Rev. Mod. Phys.}\ }\textbf {\bibinfo {volume}
  {80}},\ \bibinfo {pages} {885--964} (\bibinfo {year} {2008})}\BibitemShut
  {NoStop}%
\bibitem [{\citenamefont {Iyer}\ \emph {et~al.}(2015)\citenamefont {Iyer},
  \citenamefont {Srednicki},\ and\ \citenamefont
  {Rigol}}]{iyer_srednicki_15_103}%
  \BibitemOpen
  \bibfield  {author} {\bibinfo {author} {\bibfnamefont {D.}~\bibnamefont
  {Iyer}}, \bibinfo {author} {\bibfnamefont {M.}~\bibnamefont {Srednicki}}, \
  and\ \bibinfo {author} {\bibfnamefont {M.}~\bibnamefont {Rigol}},\ }\bibfield
   {title} {\enquote {\bibinfo {title} {Optimization of finite-size errors in
  finite-temperature calculations of unordered phases},}\ }\href {\doibase
  10.1103/PhysRevE.91.062142} {\bibfield  {journal} {\bibinfo  {journal} {Phys.
  Rev. E}\ }\textbf {\bibinfo {volume} {91}},\ \bibinfo {pages} {062142}
  (\bibinfo {year} {2015})}\BibitemShut {NoStop}%
\end{thebibliography}%

\end{document}